\documentclass[useAMS,usenatbib]{mn2e}

\RequirePackage{natbib}
\usepackage{graphicx}
\usepackage{amsmath}
\usepackage{amsfonts}
\usepackage{amssymb}
\usepackage{breqn}
\usepackage{mathtools}

\title[Dynamical Ejections of Stars due to an Accelerating Gas Filament]{Dynamical Ejections of Stars due to an Accelerating Gas Filament}

\author[T.~Boekholt, A.~Stutz, M.~Fellhauer, D.~Schleicher and D.~Matus~Carrillo]{T.~C.~N.~Boekholt$^{1,2}$\thanks{E-mail: tjardaboekholt@gmail.com (TB); astutz@astro-udec.cl (AS); mfellhauer@astro-udec.cl (MF); dschleicher@astro-udec.cl (DS) and dimatus@udec.cl (DMC)}, A.~M.~Stutz$^{1}$\footnotemark[1], M.~Fellhauer$^{1}$\footnotemark[1], D.~R.~G.~Schleicher$^{1}$\footnotemark[1] and
\newauthor D.~R.~Matus~Carrillo$^{1}$\footnotemark[1] \\
$^{1}$Departamento de Astronom\'ia, Facultad Ciencias F\'isicas y Matem\'aticas, Universidad de Concepci\'on, Av. Esteban Iturra s/n \\Barrio Universitario, Casilla 160, Concepci\'on, Chile.\\
$^{2}$CIDMA, Departamento de F\' isica, Universidade de Aveiro, Campus de Santiago, 3810-193 Aveiro, Portugal. }

\begin{document}

\date{Submitted 2017 March 31. Received 2017 March 31; in original form 2017 March 31}

\pagerange{\pageref{firstpage}--\pageref{lastpage}} \pubyear{2017}

\maketitle

\label{firstpage}

\begin{abstract}

Observations of the Orion~A integral shaped filament (ISF) have shown
indications of an oscillatory motion of the gas filament.  This
evidence is based on both the wave-like morphology of the filament as
well as the kinematics of the gas and stars, where the characteristic
velocities of the stars require a dynamical heating mechanism. As
proposed by Stutz \& Gould (2016), such a heating mechanism (the
``Slingshot'') may be the result of an oscillating gas filament in a 
gas-dominated (as opposed to stellar-mass dominated) system. Here we
test this hypothesis with the first stellar-dynamical simulations in
which the stars are subjected to the influence of an oscillating
cylindrical potential.  The accelerating, cylindrical background
potential is populated with a narrow distribution of stars.  By
coupling the potential to N-body dynamics, we are able to measure the
influence of the potential on the stellar distribution.  The
simulations provide evidence that the slingshot mechanism
can successfully reproduce several stringent observational
constraints. These include the stellar spread (both in projected
position and in velocity) around the filament, the symmetry in these
distributions, and a bulk motion of the stars with respect to the
filament.  Using simple considerations we show that star-star
interactions are incapable of reproducing these spreads on their own
when properly accounting for the gas potential.  Thus, properly
accounting for the gas potential is essential for understanding the
dynamical evolution of star forming filamentary systems in the era of
Gaia.  

\end{abstract}

\begin{keywords}
stars: kinematics and dynamics -- ISM: individual objects (Orion A)
\end{keywords}

\section{Introduction}\label{sec:intro}

Observations and simulations 
show that filamentary structures are 
abundant in the universe, e.g. they are found in the cosmic web \citep[e.g.][]{Bond96, Rieder}, but also in molecular clouds \citep[e.g.][]{Myers09, Arzou11, Contreras13, Andre14, Stutz15}. The evolution of filaments, and in particular the dynamical evolution of stellar systems 
forming inside such filamentary systems, remained mostly unexplored. In this study we present the first numerical investigation of the dynamics of young stars inside a cylindrical gravitational potential.   

We base our simulations on the well observed system in Orion A called the integral shaped filament (ISF) \citep{Bally87}.
The distribution of the young stars around this filament has the peculiar feature that the youngest stars  are mostly found on the dense, central ridgeline of the filament, while the somewhat older stars are more spread out in both projected positions and radial velocities \citep[][Fig.~13; see also Kainulainen et al.\ 2017]{SG16}\nocite{Joni17}. This can be interpreted as an evolutionary effect for which a dynamical mechanism is responsible. 

This mechanism must conform to several constraints. The dynamical heating of the stars must occur on a time scale of the order the age of the young stars, which is about 2-3\,Myr \citep{Soder14, Dunham15}. Also, it is observed that most regions along the ISF are gas mass dominated, rather than stellar mass dominated {  \citep{Stutz17}. In their Fig.~10 gravitational acceleration profiles are presented separately for the gas and stellar components. These profiles are for a slice perpendicular through the filament and through the centre of the Orion Nebula Cluster (ONC), where the stellar density is highest. There is an equality between the stellar and gas components. For slices away from the centre of the ONC however, the gas profile remains roughly the same while the stellar profile decreases, thus demonstrating the dominance of the gas component.  } As a consequence, the trajectories of the stars are primarily determined by the gas potential, rather than by the other stars, which suppresses close stellar encounters and ejections as we describe in Sec.~\ref{sec:discussion:ejec}. It is clear that the dynamical mechanism must include some kind of energy source. 

The slingshot mechanism is a new scenario proposed by \citet{SG16} in which the filament motion itself acts as the source of energy. In this scenario the stars are born in a narrow distribution along the dense, central ridgeline of a gas filament, and the distribution is broadened due to the accelerating motion of the filament itself. The integral shape of the filament in Orion A hints towards such an accelerating motion in the form of an oscillation. The origin of such an oscillation is hypothesized to be the interplay between gravitational energy and (helical) magnetic fields \citep{Heiles97}, which are observed to be of comparable strength \citep{SG16}. 
In particular, magnetic tension can give rise to a restoring force after an initial bending of the filament, to which also gravity can contribute { \citep{Peters12, Schleicher17}. }
While an analysis for massive filaments such as the ISF is still missing, the possibility to drive oscillations in low-mass filaments by gravitational interactions has been shown by \citet{Grit17}.
In our study we will assume that the oscillations are driven by some external mechanism, and focus on the kinematics of stars under the influence of such an oscillating, cylindrical potential.   

The acceleration of a gas filament has profound consequences for any stellar distribution located inside the filament. This study presents the first test of the slingshot mechanism as the main mechanism responsible for the dynamical heating of the stars. We perform numerical simulations of the dynamics and kinematics of stars inside an accelerating,  cylindrical potential and give numerical evidence that stars can get kinematically heated by the filament. This dynamical heating is fully determined by the acceleration of the gas filament's motion. We will also show that a symmetric broadening of the phase space distribution of the stars occurs naturally.

We use the AMUSE framework { \citep{2009NewA...14..369P, PortegiesZwart2013456, AMUSE13} } to couple N-body dynamics to the gravitational potential of the gas filament as measured by \citet{SG16} (see Sec.~\ref{sec:method}). We populate the filament with stars and simulate the effect of an oscillating potential, with varying oscillation amplitude and period. We confirm that an accelerating filament can increase the spread of stars around the filament (see Sec.~\ref{sec:results}). This positive result motivates subsequent studies, including the evolution of a star cluster in a moving filament, and the actual decoupling mechanism of stars from the gas filament. We finish by discussing our main findings (see Sec.~\ref{sec:discussion}) and summing up our main conclusions (see Sec.~\ref{sec:conclusion}).   

\section{Method}\label{sec:method}

In this section we present our model of the gas filament, which is approximated by a cylindrical, background potential. We describe the main ingredients of our simulation code and perform a validation test. 

\subsection{The gas filament}

\begin{figure*}
\centering
\begin{tabular}{ccc}
\includegraphics[height=0.26\textwidth,width=0.3\textwidth]{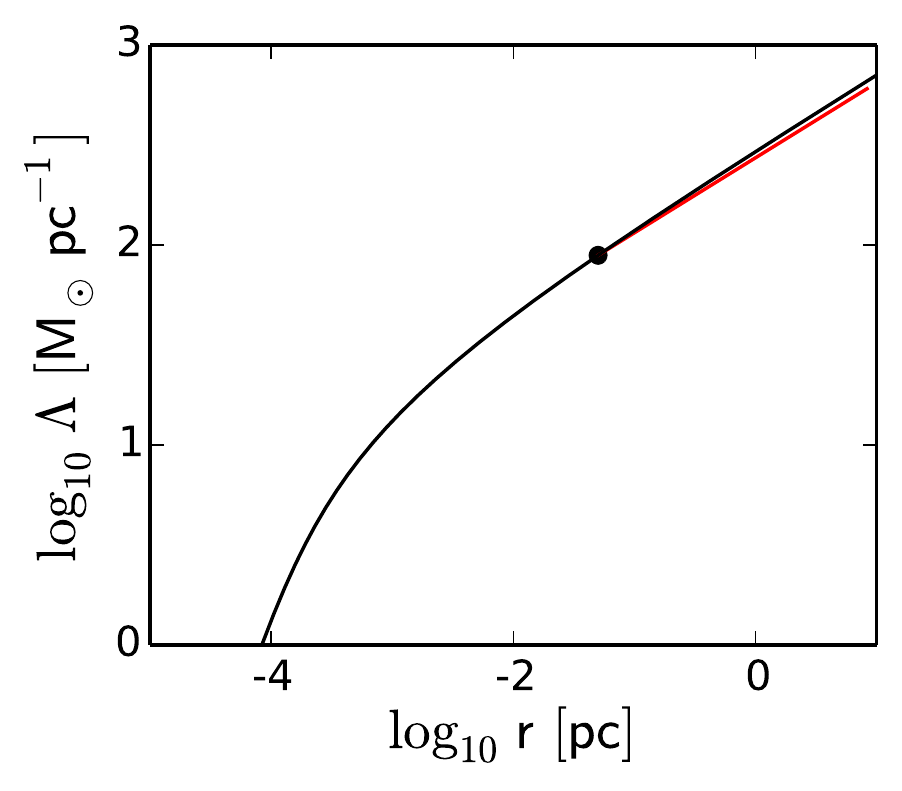} &
\includegraphics[height=0.26\textwidth,width=0.3\textwidth]{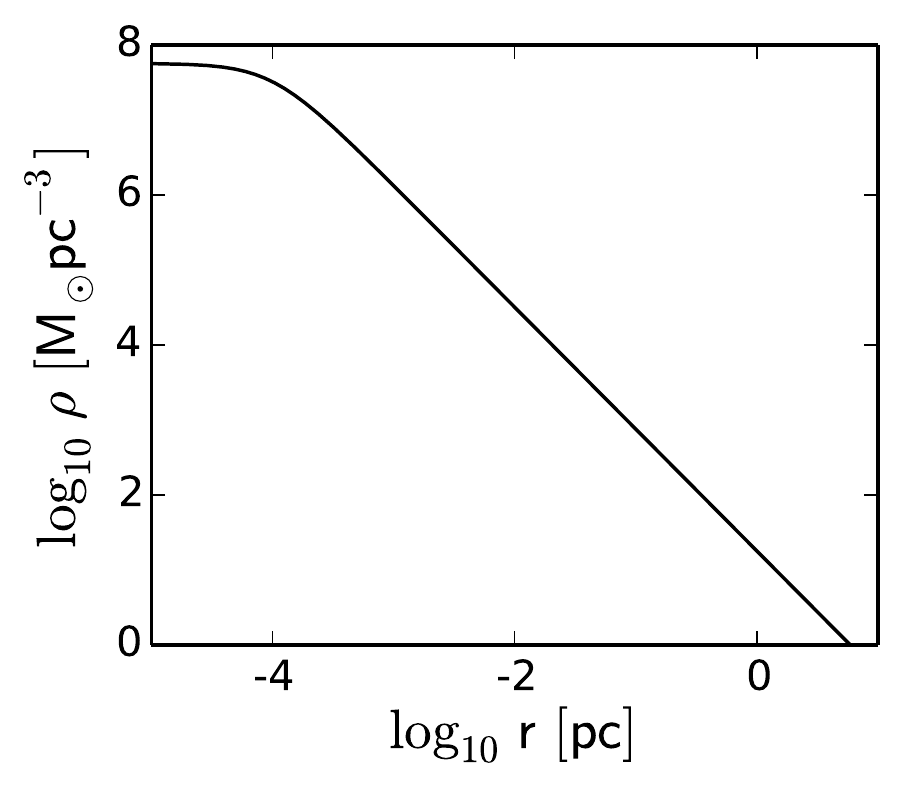} &
\includegraphics[height=0.26\textwidth,width=0.3\textwidth]{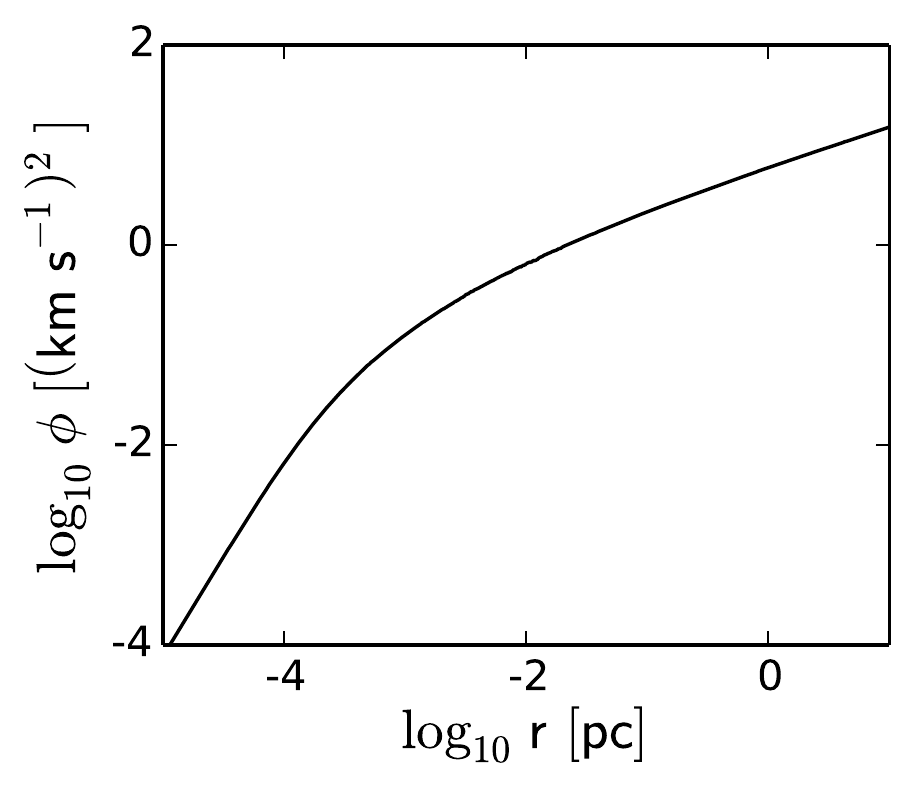} \\
\end{tabular}  
\caption{ Model of the gas filament: line density (left), volume density (middle) and potential (right). We adopt a softening parameter $D=10^{-4}$\,pc. Such a small value is necessary in order to reproduce the straightness of the line density profile (see red curve in left panel which represents the model of \citet{SG16}). We fit their model at $\Lambda \left( 0.05\,\rm{pc} \right) = 89\,\rm{M_\odot\,pc^{-1}}$ (represented by the dot).   }
\label{fig:model}
\end{figure*}

We base our model of the gas filament on that of \citet{SG16}, who construct a cylindrical potential: $\phi \left( r \right) = \phi_0 r^{\gamma}$, with $r$ the cylindrical radius, $\phi_0\rm{ = 6.3\,(\rm{km\,s^{-1})^2}}$ and $\gamma = 3/8$. While the potential is well behaved at $r=0$, the gravitational acceleration and density diverge. { To resolve this issue and to make the model more physical, we flatten the density profile by introducing a polytrope-like softening.} The expressions for the volume density, line density, and related gravitational acceleration are respectively:

\begin{equation}
{{\rho \left( r \right)}} = \rho_0 \left( 1 + \left( {r \over D} \right)^2 \right)^{{\gamma - 2} \over 2}
\label{eq:density}
\end{equation}

\begin{equation}
\Lambda \left( r \right) = \Lambda_0 \left( \left( 1 + \left( r / D \right)^2 \right)^{\gamma \over 2} - 1 \right) 
\label{eq:linemass}
\end{equation}

\begin{equation}
{\vec{g}\left( r \right)} =
\begin{cases}
- {2 G \Lambda \left( r \right) \over r}~\hat{r}, & r > 0\\
0~\hat{r} & r = 0.
\end{cases}
\label{eq:grav}
\end{equation}

\noindent Here $\rho_0$ is the density in the flattened region, $D$ is the polytrope-like softening radius, $\gamma$ is the power law index of the gas potential, and $\Lambda_0 = 2 \pi \rho_0 {D^2 \over \gamma}$, which is obtained by integrating the volume density in cylindrical coordinates. The related gravitational potential cannot be expressed in a simple equation and we therefore solve for the potential through numerical integration. For large radii, e.g. $r \gg D$, our profiles converge to those of \citet{SG16}. The line density, volume density, and potential profiles are shown in Fig.~\ref{fig:model}. 

The main observational constraint on our model of the gas filament is the line mass profile \citep[Fig.~5]{SG16}, e.g. the integrated mass as a function of position along the filament. Assuming a cylindrical geometry, they derive the line density in their Eq.~6, which we plot as the red curve in Fig.~\ref{fig:model} (left panel). The derived line density is very accurately modelled by a power law from $r=0.05$\,pc to $r=8.5$\,pc. In order for our polytrope-like model to be consistent with this profile, we fit our model to the line density value of $\Lambda \left( 0.05\,\rm{pc} \right) = 89\,\rm{M_\odot\,pc^{-1}}$ (see the dot in Fig.~\ref{fig:model}, left panel), and set the value of the softening parameter to $D = 10^{-4}$\,pc. For larger values of $D$, there will be an excess of curvature in the line density profile within the observed range in radius, which is caused by the flattening of the density profile. By adopting the value of $D=10^{-4}$\,pc we limit the error in the line density at $r=8.5$\,pc to less than 10 percent. Although the small value of $D$ was somewhat unexpected, recent simulations show that power law density profiles of filaments can persist up to very small radii if the filament is gravity dominated \citep[Fig.~17]{Clarke17}. 

{ The regime of validity of the presented model is confined to the inner few parsecs from the filament centre. The gas profile, based on observations and giving $\gamma=3/8$, is a profile that gives infinite mass when integrated to infinity. Therefore any realistic model should have a truncation radius after which the density steeply drops. The observations from \citet{SG16} go out to $\sim$ 8.5\,pc and do not show such a cut-off, implying that this should occur at even larger radius. Since most of our stellar interactions occur within 8.5\,pc and since we are mostly interested in the stellar distribution within this similar range, the cut-off in the density profile will be of limited importance to our results. It is rather the enclosed mass that governs the dynamics of the stars. }

Throughout this study we will assume that our gas potential is oscillating in a radial direction. The origin of the filament moves along the x-axis in a sinusoidal fashion: 
\begin{equation}
x_0\left( t \right) = A \sin \left( {2 \pi \over P} t \right),
\end{equation}  
\noindent with $x_0$ the centre of the potential, $A$ the amplitude and $P$ the period of the oscillation.
Later in this study we will refer to the maximum velocity and acceleration, which are respectively: $v_{\rm{max}} = 2 \pi A/P$ and $a_{\rm{max}} = 4 \pi^2 A / P^2$. For small values of $A$ and large values of $P$ our model  reduces to a static filament. 
{ Even though in reality the filament's motion might be more complex, a (mathematical) cylinder of infinite extent that performs simple harmonic motion is the simplest model that can be analysed in detail. This model neglects clumps in the gas mass distribution, a finite length of the cylinder and a more complex and realistic filament motion. Our main aim is to illustrate the slingshot effect and provide an anchor point for further studies introducing improvements to the model.}

\subsection{\texttt{AMUSE} and \texttt{BRIDGE}}

The Astrophysical Multi-purpose Software Environment (\texttt{AMUSE}) { \citep{2009NewA...14..369P, PortegiesZwart2013456, AMUSE13} } is a user friendly software framework for coupling different physics codes into one self-consistent simulation. For this study we require the coupling of N-body dynamics and a time-varying gas potential. Several previous studies have successfully coupled N-body dynamics to a background potential using \texttt{AMUSE}, see for example \citet{siblings16}. \texttt{AMUSE} also offers great flexibility to the user in implementing a new gas potential profile, varying its parameters, and introducing a time-dependence. Therefore, we choose \texttt{AMUSE} to perform our simulations. 

\texttt{AMUSE} also provides a method for coupling an N-body code to a background potential in a modular fashion via the \texttt{BRIDGE} method. For a comprehensive explanation of this method we refer the reader to \citet{bridge07}. In short, the velocities of the stars will regularly get updated taking into account the acceleration due to the gas filament, while in between these updates, the stellar kinematics are evolved as if the stars were in isolation to account for their respective interactions. In order for this algorithm to be accurate, it is important to use a sufficiently small \texttt{BRIDGE} time step size. For our simulations we find empirically that a time step size of a 100 years gives sufficient accuracy.
We implemented the gas model defined in the previous section into \texttt{AMUSE}, and use the \texttt{HUAYNO} N-body code \citep{huayno12} to evolve the stars in isolation. 

\subsection{Validation}\label{Sec:method:validation}

\begin{figure}
\centering
\begin{tabular}{cc}
\includegraphics[height=0.32\textwidth,width=0.4\textwidth]{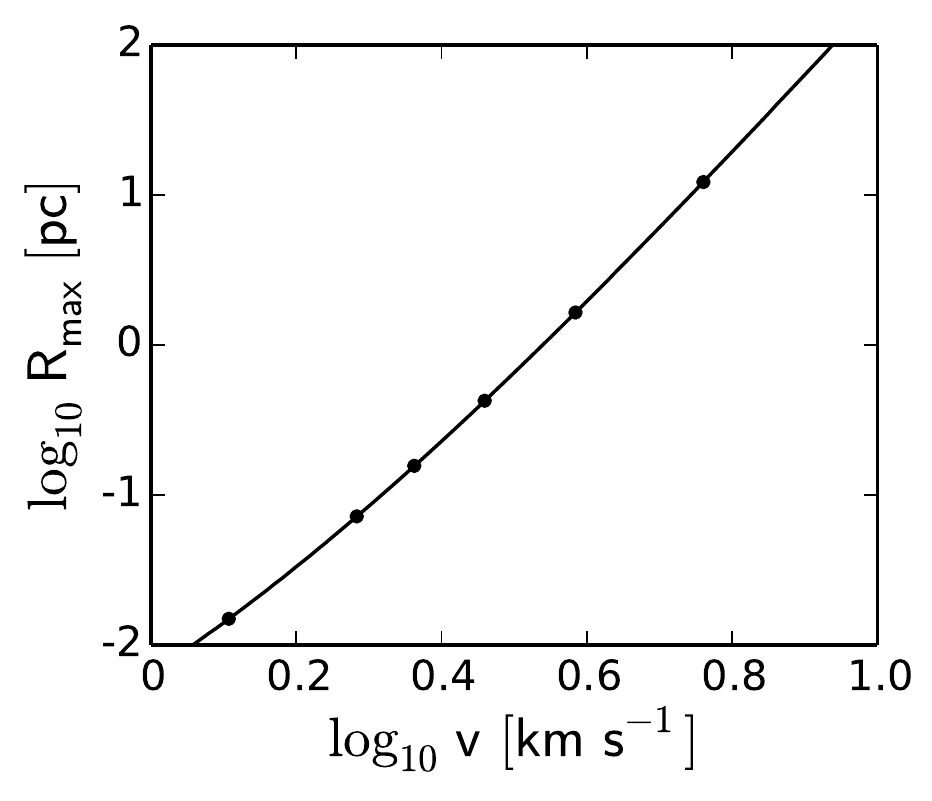} \\
\end{tabular}  
\caption{ Maximum radial distance from the filament versus initial ejection velocity, for a star at the centre of the filament. The numerical results (data points) are consistent with the analytical prediction (curve).  }
\label{fig:validation}
\end{figure}

In order to validate our \texttt{AMUSE} script, we perform a stellar ejection experiment and compare the result to the analytical prediction. 
The initial configuration of our numerical experiment consists of our model for the gas potential and a single star placed in its centre. We give both the filament and the star the same initial velocity, so they are at rest with respect to each other. The filament is not oscillating in this experiment, but is in uniform motion until it is forced to stop instantaneously at a distance of 2.5\,pc from its starting point. As a result, the star will get ejected out of the centre and reach a certain maximum distance from the filament, which we define as $R_{\rm{max}}$. Note that in our model there is no escape velocity, as we have not introduced a truncation radius. We vary the initial velocity, $v$, of the filament and measure $R_{\rm{max}}$. The result is given in Fig.~\ref{fig:validation} (see data points).  

The value of $R_{\rm{max}}$ as a function of initial velocity can be calculated analytically by solving the equation $\phi\left( R_{\rm{max}} \right) = {1 \over 2} v^2 + \phi\left( r \right)$, where $r$ and $v$ are the initial displacement and velocity respectively. The value for $R_{\rm{max}}$ is obtained by first evaluating $\phi \left( r \right)$ using numerical integration and then integrating $g \left( r \right)$ until the integral equals ${1 \over 2} v^2 + \phi \left( r \right)$. For a zero initial displacement, we show the analytical result in Fig.~\ref{fig:validation} (solid curve).   

We confirm that the simulation data perfectly match the analytical prediction. This result provides a robust anchor point for the reliability of our subsequent experiments, in which we study the dynamical interaction of a larger number of stars with an oscillating gas potential. 

\section{Experimental setup}

\begin{table*}
\centering
\begin{tabular}{ccccccccccc}
\hline
N & $\sigma_{\rm{r}}$ & A & P & $a_{\rm{max}}$ & T & $N_{\rm{runs}}$ & $f_{\rm{filament}}$ & $x_{\rm{av}}$ & $v_{\rm{av}} $ & $\lambda$\\
 & [pc] & [pc] & [Myr] & [$\rm{km\,s^{-1}\,Myr^{-1}}$] & [Myr] &  &  & [pc] & [$\rm{km\,s^{-1}}$] & [$\rm{Myr^{-1}}$] \\
\hline 			
400 & 0.05 & 0.5   & 2.5  & 3.16 & 5.0 & 1 & 0.87 & 0.07 & 0.70 & 0.02 \\
400 & 0.05 & 0.5   & 2.25 & 3.90 & 5.0 & 1 & 0.83 & 0.14 & 0.68 & 0.04  \\
400 & 0.05 & 0.5   & 2.0  & 4.93 & 5.0 & 1 & 0.81 & 0.20 & 0.79 & 0.05  \\
400 & 0.05 & 0.5   & 1.75 & 6.45 & 5.0 & 1 & 0.72 & 0.38 & 0.75 & 0.06  \\
400 & 0.05 & 0.5   & 1.5  & 8.77 & 5.0 & 1 & 0.51 & 0.53 & 1.11 & 0.10  \\
400 & 0.05 & 0.5   & 1.25 & 12.6 & 5.0 & 1 & 0.25 & 1.18 & 2.11 & 0.16  \\
400 & 0.05 & 0.5   & 1.0  & 19.7 & 5.0 & 1 & 0.10 & 1.62 & 2.84 & 0.18  \\
400 & 0.05 & 0.5   & 0.75 & 35.1 & 5.0 & 1 & 0.03 & 2.76 & 2.15 & 0.19  \\
\hline
400 & 0.05 & 0.75  & 2.5  & 4.74 & 5.0 & 1 & 0.86 & 0.14 & 0.68 & 0.03  \\
400 & 0.05 & 0.75  & 2.25 & 5.85 & 5.0 & 1 & 0.78 & 0.37 & 0.78 & 0.05  \\
400 & 0.05 & 0.75  & 2.0  & 7.40 & 5.0 & 1 & 0.68 & 0.72 & 1.26 & 0.07  \\
400 & 0.05 & 0.75  & 1.75 & 9.67 & 5.0 & 1 & 0.49 & 1.15 & 1.00 & 0.11  \\
400 & 0.05 & 0.75  & 1.5  & 13.2 & 5.0 & 1 & 0.26 & 1.83 & 1.91 & 0.16  \\
400 & 0.05 & 0.75  & 1.25 & 18.9 & 5.0 & 1 & 0.11 & 3.03 & 3.63 & 0.20  \\
400 & 0.05 & 0.75  & 1.0  & 29.6 & 5.0 & 1 & 0.04 & 2.91 & 4.89 & 0.23  \\
\hline
400 & 0.05 & 1.0   & 2.5  & 6.32 & 5.0 & 1 & 0.75 & 0.57 & 0.99 & 0.06  \\
400 & 0.05 & 1.0   & 2.25 & 7.80 & 5.0 & 1 & 0.62 & 1.31 & 0.77 & 0.09  \\
400 & 0.05 & 1.0   & 2.0  & 9.87 & 5.0 & 1 & 0.46 & 1.74 & 2.16 & 0.14  \\
400 & 0.05 & 1.0   & 1.75 & 12.9 & 5.0 & 1 & 0.27 & 1.98 & 1.58 & 0.10  \\
400 & 0.05 & 1.0   & 1.5  & 17.5 & 5.0 & 1 & 0.17 & 2.55 & 2.05 & 0.15  \\
400 & 0.05 & 1.0   & 1.25 & 25.3 & 5.0 & 1 & 0.08 & 2.32 & 5.65 & 0.19  \\
\hline
\end{tabular}
\caption{ Overview of the simulations concerning Fig.~\ref{fig:exp_decay}. The first four parameters describe the initial conditions: $\rm{N}$ = number of stars, $\sigma_r$ = Gaussian spread of initial displacements, $\rm{A}$ = oscillation amplitude and $\rm{P}$ = oscillation period. Next three parameters are: $a_{\rm{max}}$ = maximum acceleration of the filament, $\rm{T}$ = simulation time and $N_{\rm{runs}}$ = number of runs. The last four quantities describe the stellar distribution at the end of the simulation: $f_{\rm{filament}}$ = fraction of stars still on the filament, $x_{\rm{av}}$ = average absolute distance to the filament, $v_{\rm{av}}$ = average absolute relative velocity with respect to the filament, and $\lambda$ is the fitted exponent of the approximate exponential decay of the stellar distribution on the filament.}
\label{tab:simulations}
\end{table*}

\begin{table*}
\centering
\begin{tabular}{ccccccccccccccc}
\hline
N & $\sigma_r$ & A & P & $a_{\rm{max}}$ & T & $N_{\rm{runs}}$ & $\rm{f_l}$ & $\rm{f_r}$ & $x_{\rm{av,l}}$ & $x_{\rm{av,r}}$ & ${\sigma_{\rm{r,l}}}$ & ${\sigma_{\rm{r,r}}}$ & $vx_{\rm{av,l}}$ & $vx_{\rm{av,r}}$ \\
 & [pc] & [pc] & [Myr] & [$\rm{km\,s^{-1}\,Myr^{-1}}$] & [Myr] &  & & & [pc] & [pc] & [pc] & [pc] & [$\rm{km\,s^{-1}}$] & [$\rm{km\,s^{-1}}$] \\
\hline 			
128 & 0.05 & 0.5   & 2.0 & 4.93 & 2.0 & 10 & 0.06 & 0.02 & 0.77 & 0.06 & 0.82 & 0.01 & 1.03 & 1.40 \\
128 & 0.05 & 1.0   & 2.0 & 9.87 & 2.0 & 10 & 0.27 & 0.10 & 1.65 & 0.97 & 1.12 & 0.92 & 1.33 & 1.75 \\
128 & 0.05 & 1.25  & 2.0 & 12.3 & 2.0 & 10 & 0.25 & 0.29 & 2.29 & 2.03 & 1.20 & 1.19 & 2.19 & 0.88 \\
128 & 0.05 & 1.5   & 2.0 & 14.8 & 2.0 & 10 & 0.26 & 0.42 & 1.86 & 3.51 & 1.39 & 1.33 & 2.56 & 0.11 \\
128 & 0.05 & 2.0   & 2.0 & 19.7 & 2.0 & 10 & 0.12 & 0.67 & 2.43 & 5.80 & 1.84 & 2.38 & 3.21 & 1.39 \\
128 & 0.05 & 2.5   & 2.0 & 24.7 & 2.0 & 10 & 0.05 & 0.82 & 2.40 & 8.73 & 1.71 & 3.12 & 2.66 & 2.98 \\
\hline
\end{tabular}
\caption{ Overview of the simulations concerning Fig.~\ref{fig:symmetry}. The first seven parameters are similar to those of Tab.~\ref{tab:simulations}. The last eight quantities describe the symmetry of the stellar distribution at the end of the simulation: $\rm{f}$ is the fraction of the total number of stars, $x_{\rm{av}}$ is the average distance to the filament, $\sigma_{\rm{r}}$ is the spread around $x_{\rm{av}}$, and $vx_{\rm{av}}$ is the average relative velocity with respect to the filament. We denote the stars to the left or right of the filament with subscripts $l$ and $r$ respectively.}
\label{tab:simulations2}
\end{table*}

The accelerating, cylindrical potential introduced in the previous section will represent the gas filament. Its oscillation is specified by two parameters: $A$ and $P$, which are the oscillation amplitude and the oscillation period respectively. We perform a grid study of these two parameters in order to quantify the relation between stellar ejections and filament motion, and to evaluate how stellar ejections depend on the acceleration of the filament.  

In this study we adopt equal mass stars with $\rm{M = 0.5}$\,$\rm{M_\odot}$.
{ This value corresponds to the peak value of a Kroupa initial mass function. The effect of a mass spectrum in the context of accreting protostars in an oscillating filament is left for a follow-up study. }
We give the stars a small initial displacement from the centre of the filament, but zero initial velocity.
{ Since we constrain the filament's motion to one dimension, the most important velocity component of the stars concerning the slingshot ejection will be the component parallel to the filament motion. Our cold initial conditions successfully model a distribution of parallel velocity components with a certain velocity dispersion.}

The stellar initial condition is fully specified by two parameters: $N$, the number of stars along the 16\,pc filament, and $\sigma_r$, the Gaussian standard deviation of the initial displacements. We set $\sigma_r = 0.05$\,pc, which is the resolution limit of the observations analysed by \citet{SG16}.  
Inside $r = 0.05$\,pc we have an enclosed gas mass $\sim 1500\,M_\odot$. In our simulations this mass scale sets the approximate transition from a gas to stellar mass dominated system. We adopt a number of stars $N \lesssim 10^3$ such that the system is gas mass dominated, consistent with the observations \citep{Stutz17}. 
As a result, each star will oscillate through the centre of the filament following the gas potential.
In addition there will be small gravitational perturbations due to the other stars.
Conceptually, this configuration will approximately mimic a filament where all the stars are ``born'' at the same time and are distributed along the filament ridgeline. A complete overview of our simulation input and output parameters is provided in Tab.~\ref{tab:simulations} and \ref{tab:simulations2}.

\section{Results}\label{sec:results}

\begin{figure*}
\centering
\begin{tabular}{ccccc}
\includegraphics[height=0.2\textwidth,width=0.2\textwidth]{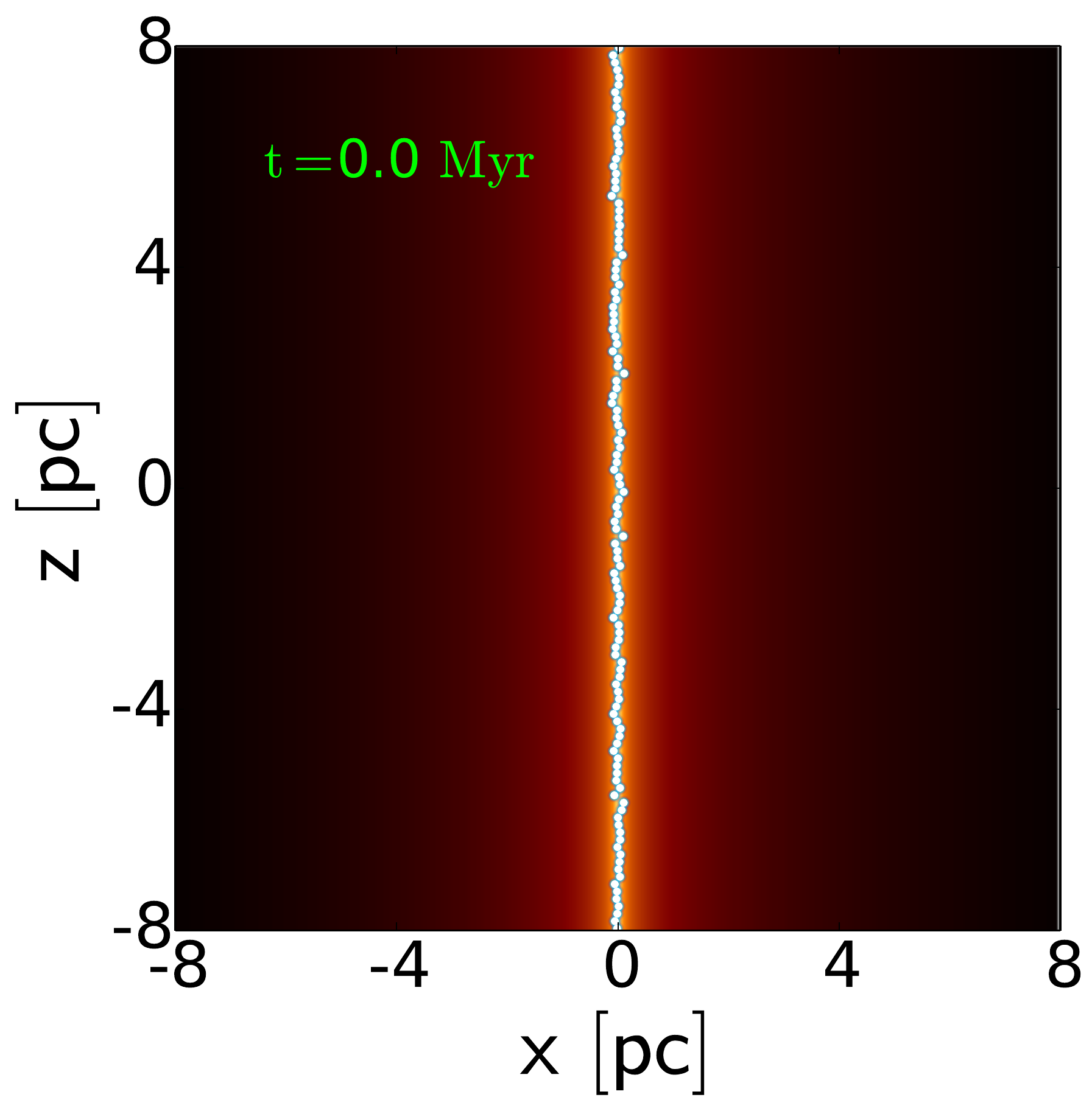} &
\hspace{-4mm}
\includegraphics[height=0.2\textwidth,width=0.2\textwidth]{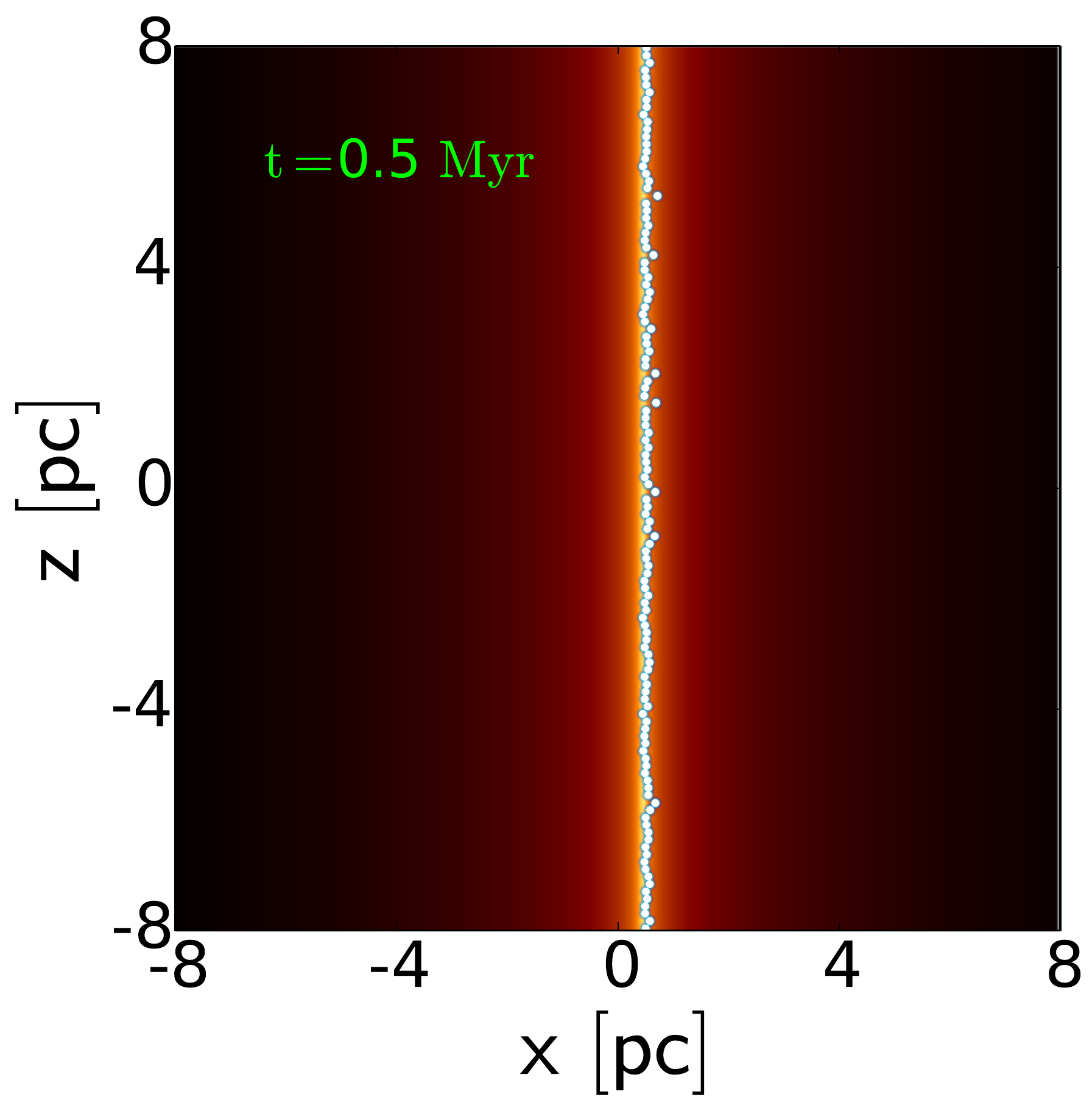}  &
\hspace{-4mm}
\includegraphics[height=0.2\textwidth,width=0.2\textwidth]{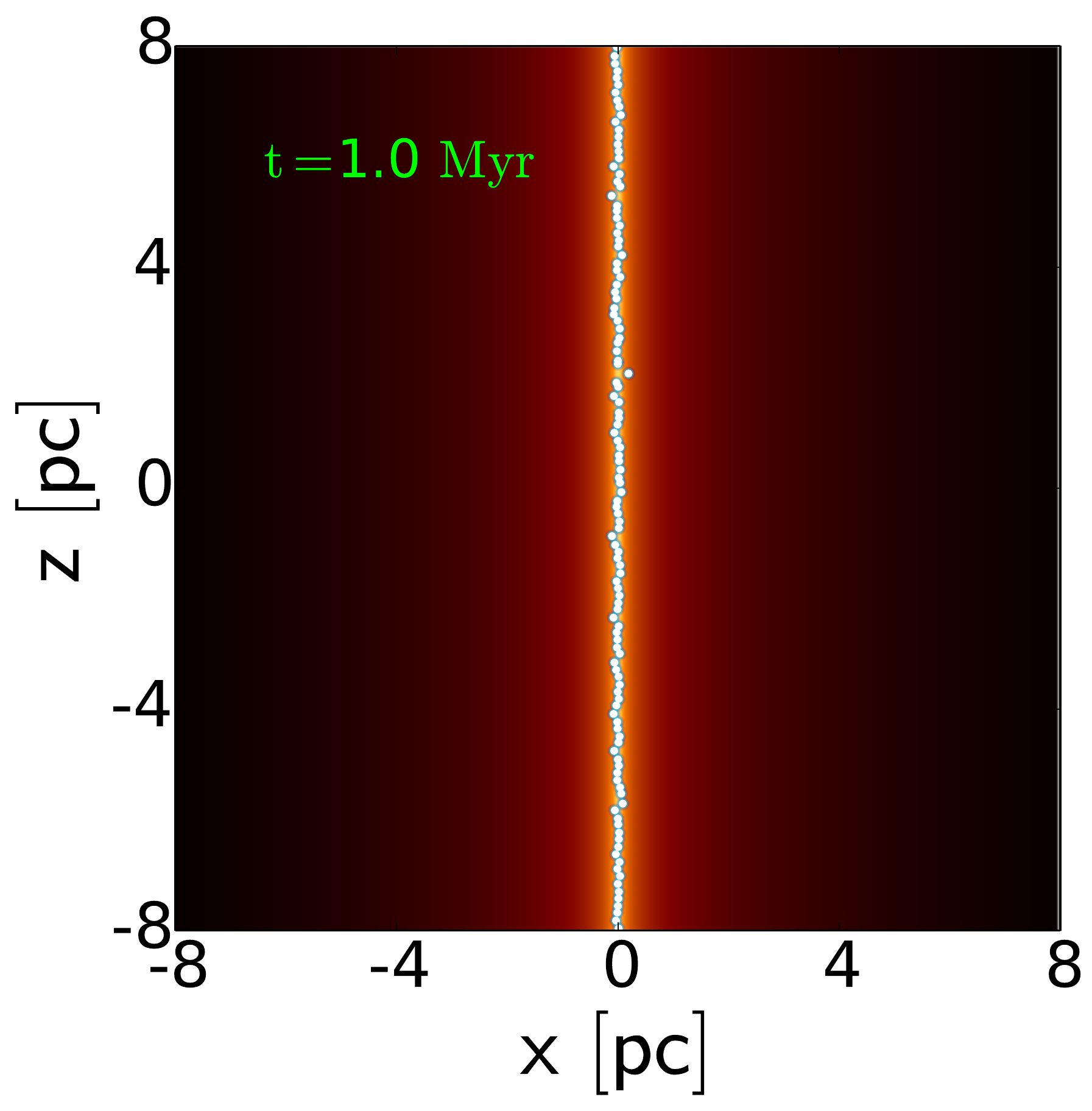} &
\hspace{-4mm}
\includegraphics[height=0.2\textwidth,width=0.2\textwidth]{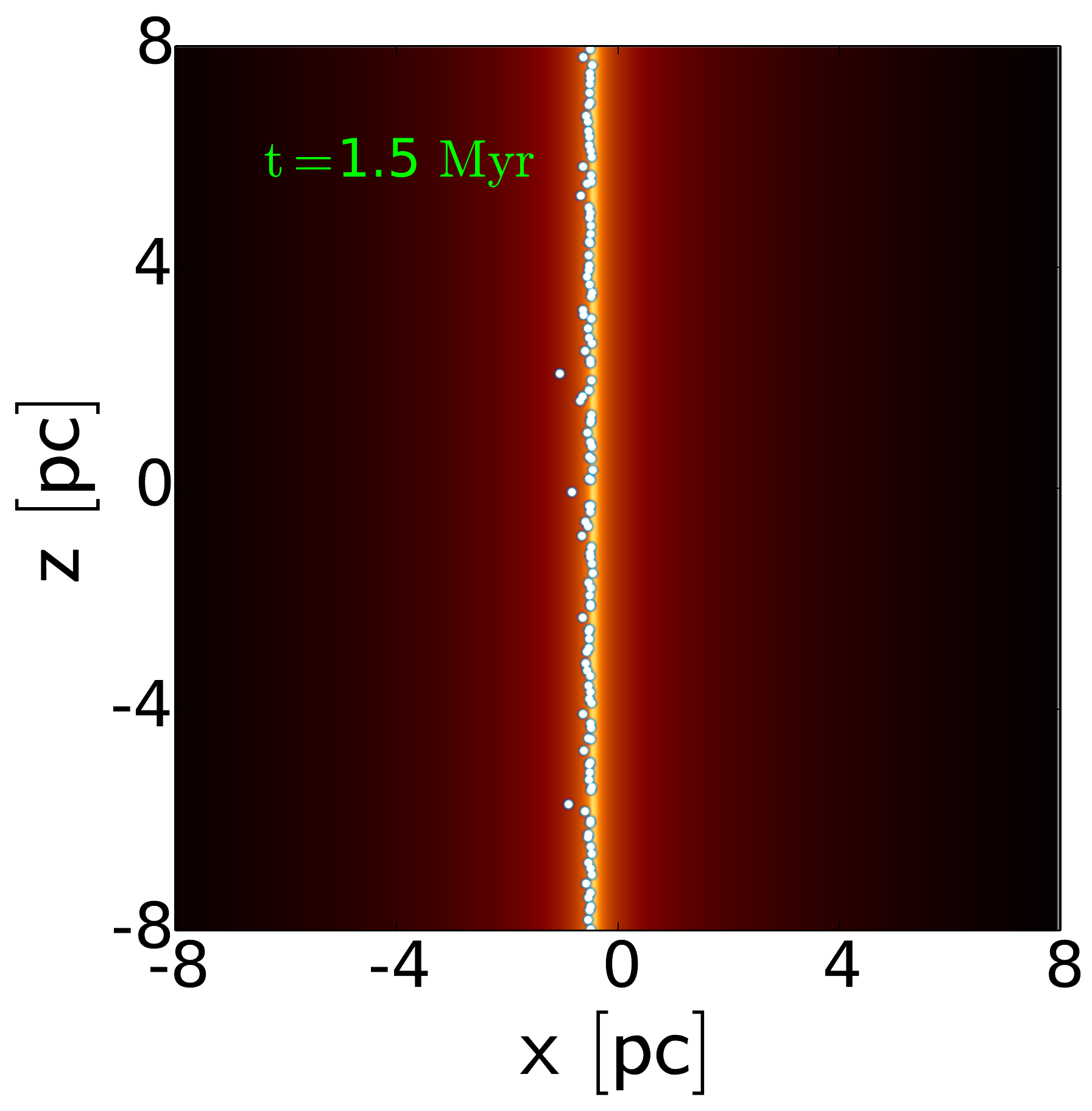} &
\hspace{-4mm}
\includegraphics[height=0.2\textwidth,width=0.2\textwidth]{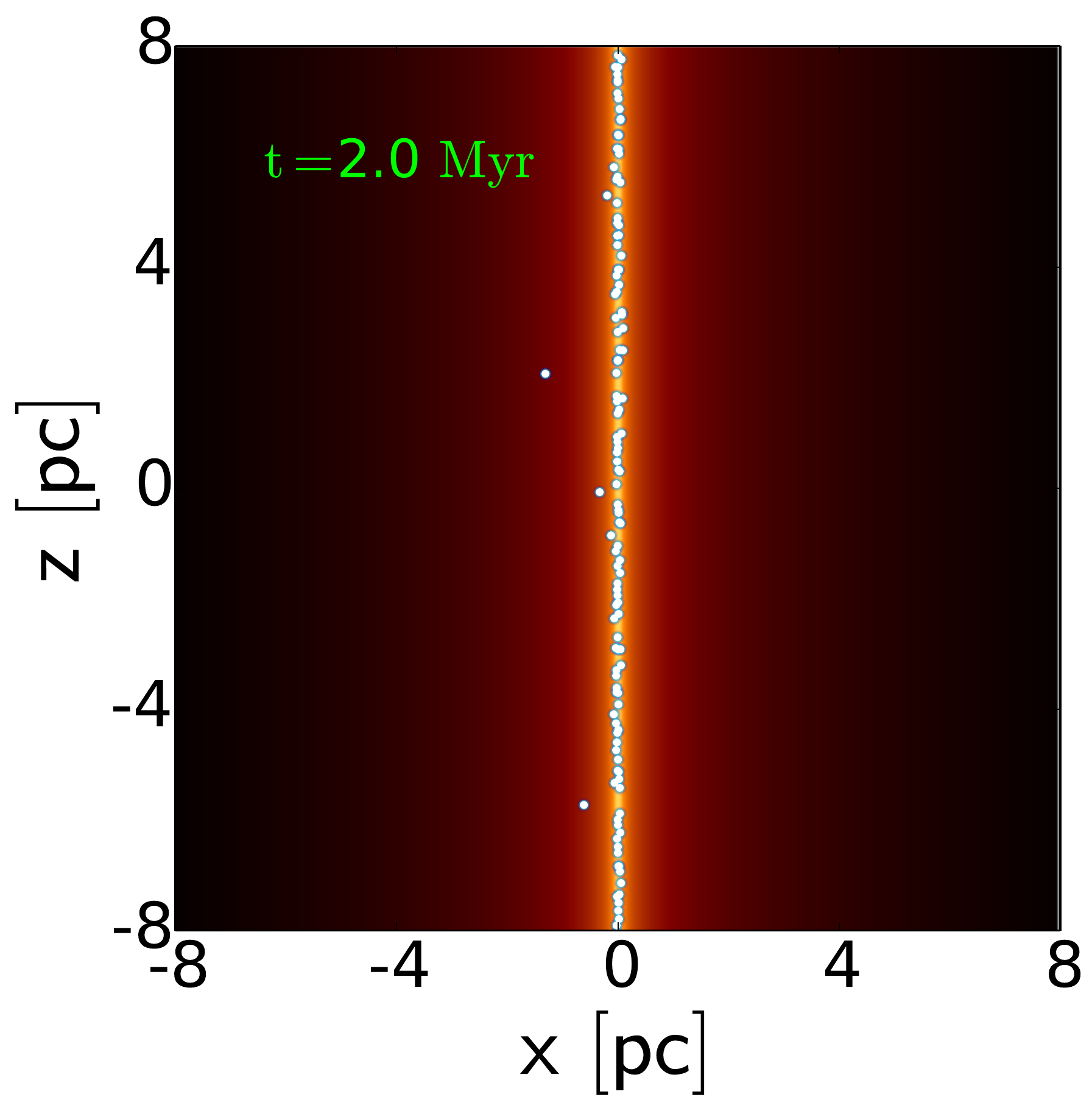}  \\

\includegraphics[height=0.2\textwidth,width=0.2\textwidth]{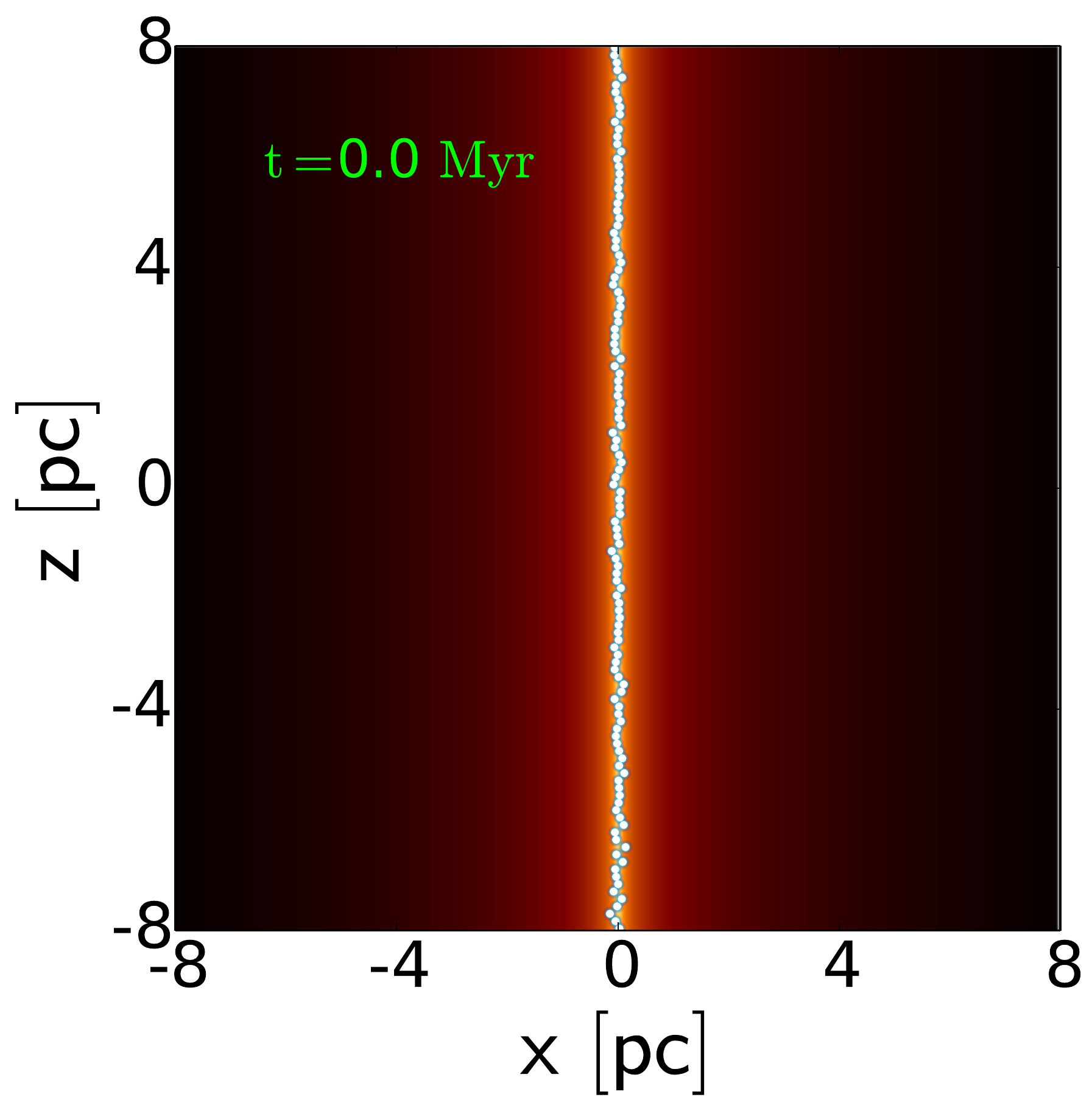} &
\hspace{-4mm}
\includegraphics[height=0.2\textwidth,width=0.2\textwidth]{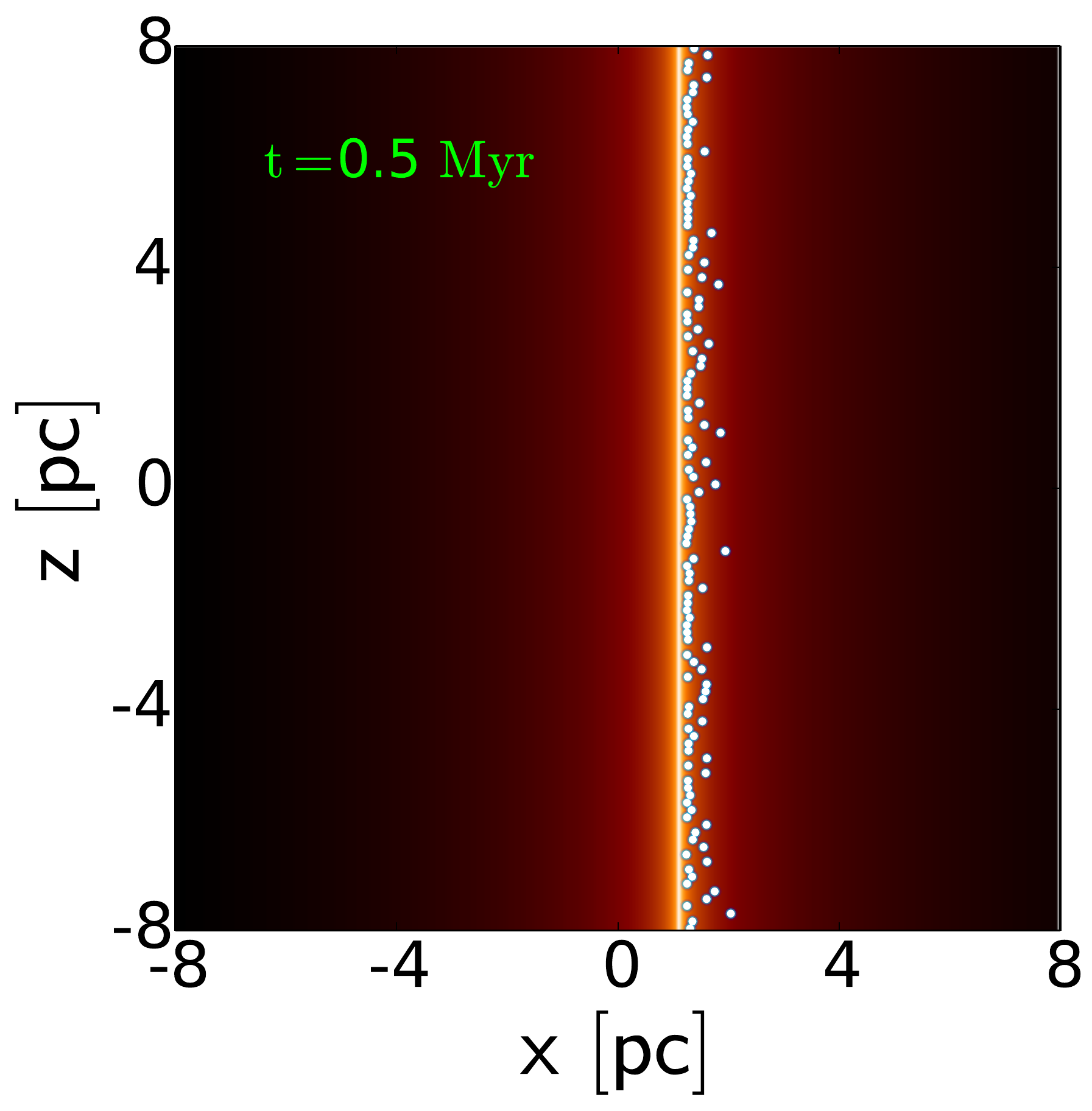}  &
\hspace{-4mm}
\includegraphics[height=0.2\textwidth,width=0.2\textwidth]{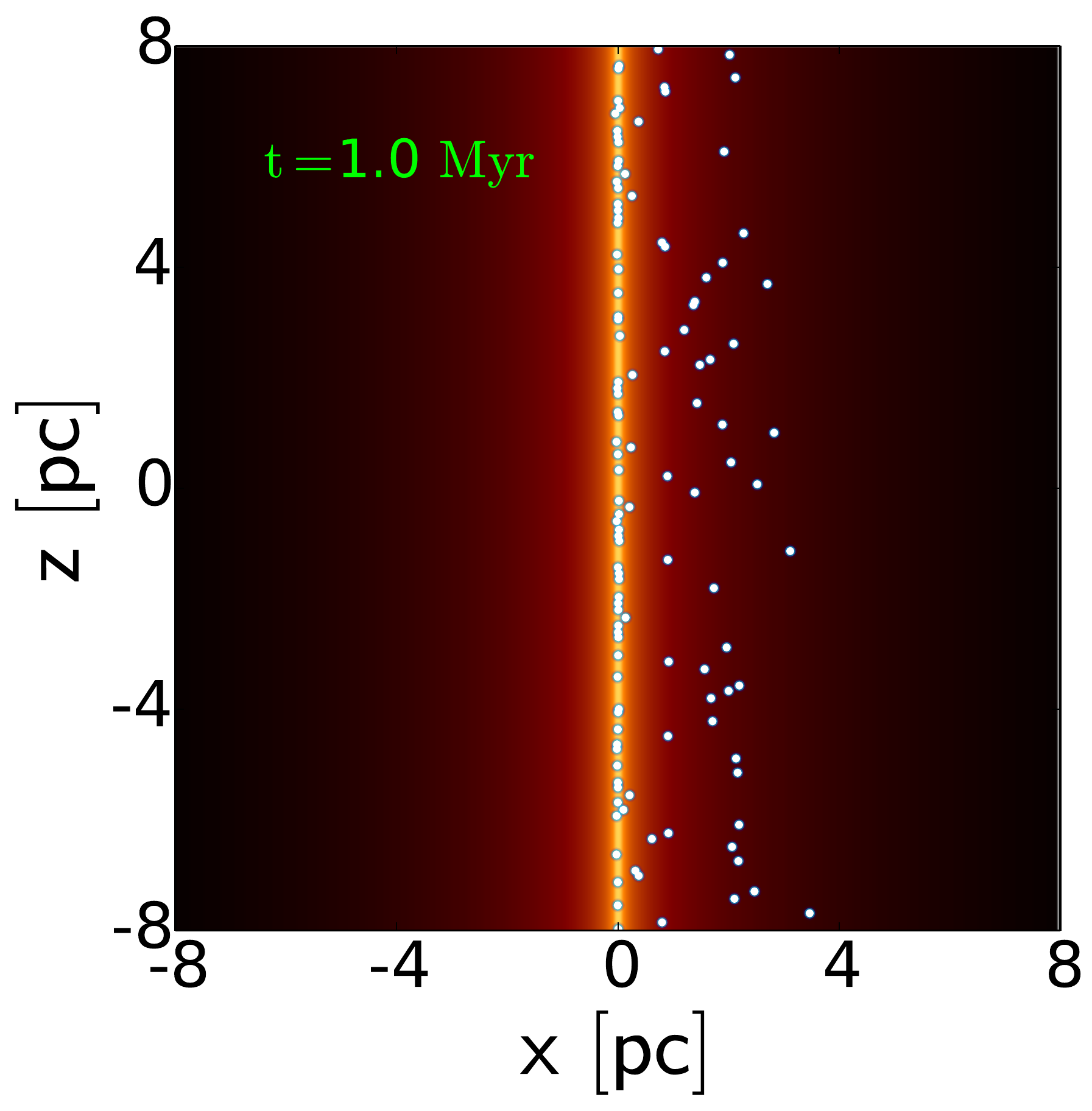} &
\hspace{-4mm}
\includegraphics[height=0.2\textwidth,width=0.2\textwidth]{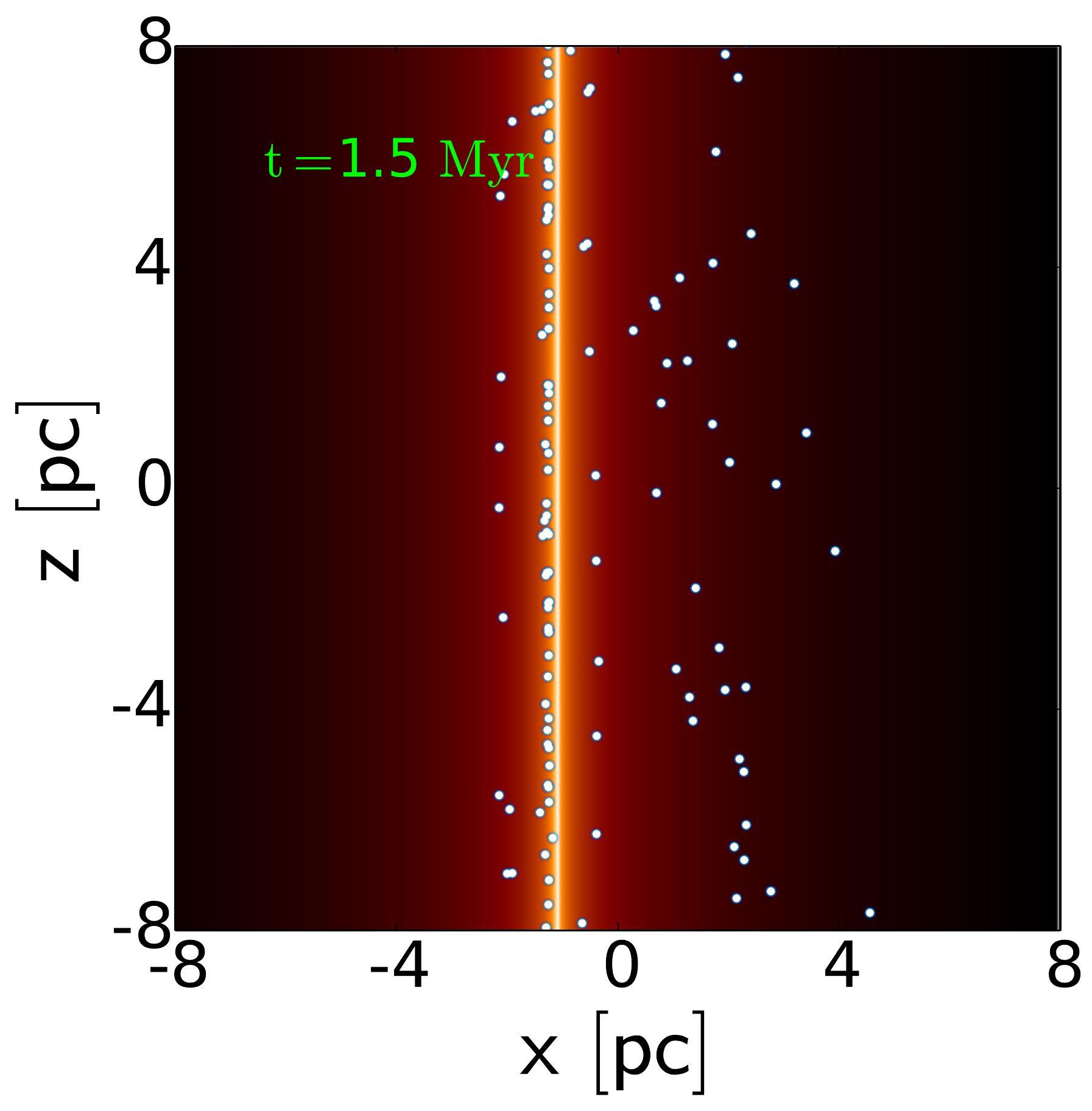} &
\hspace{-4mm}
\includegraphics[height=0.2\textwidth,width=0.2\textwidth]{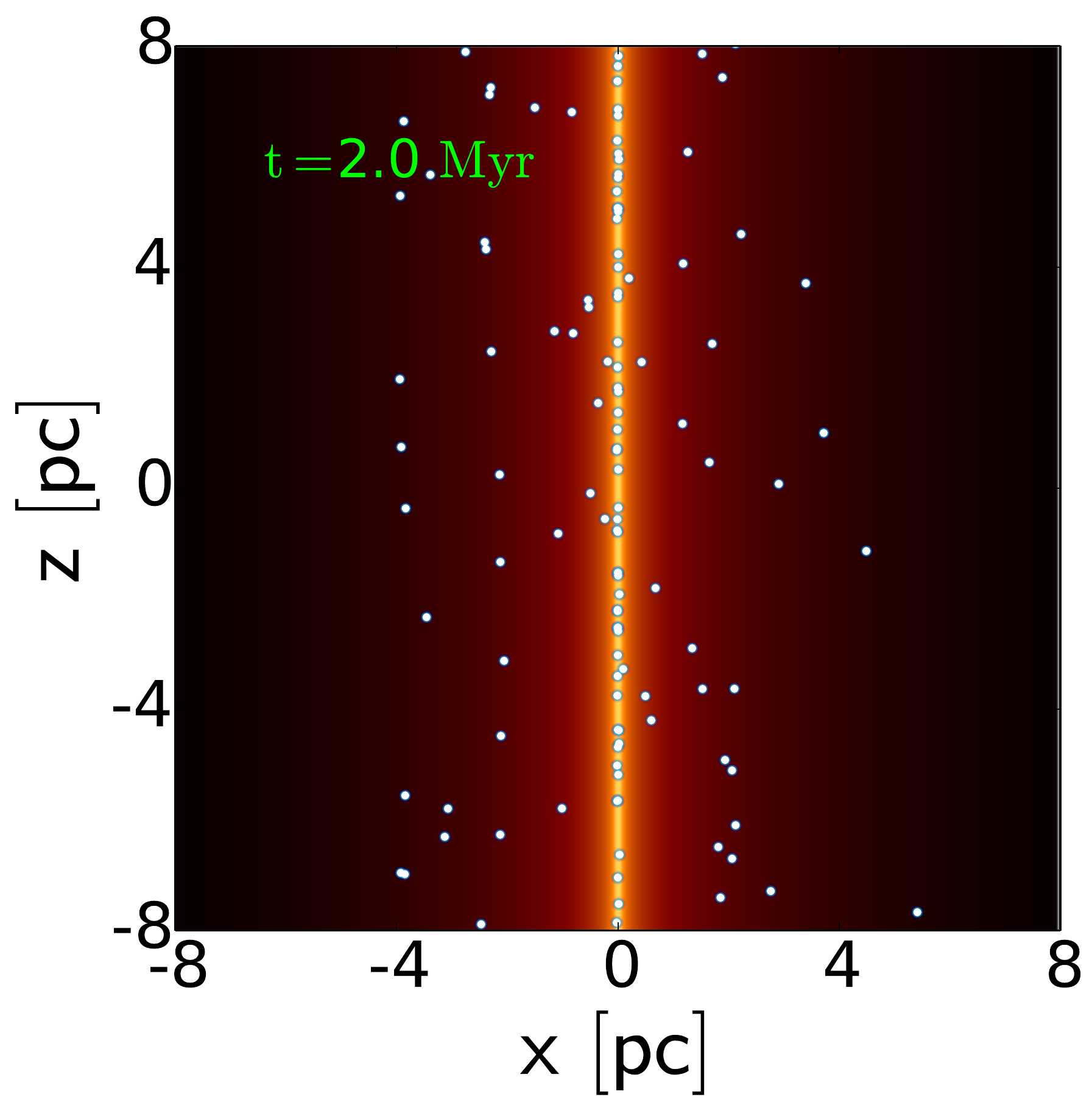}  \\

\includegraphics[height=0.2\textwidth,width=0.2\textwidth]{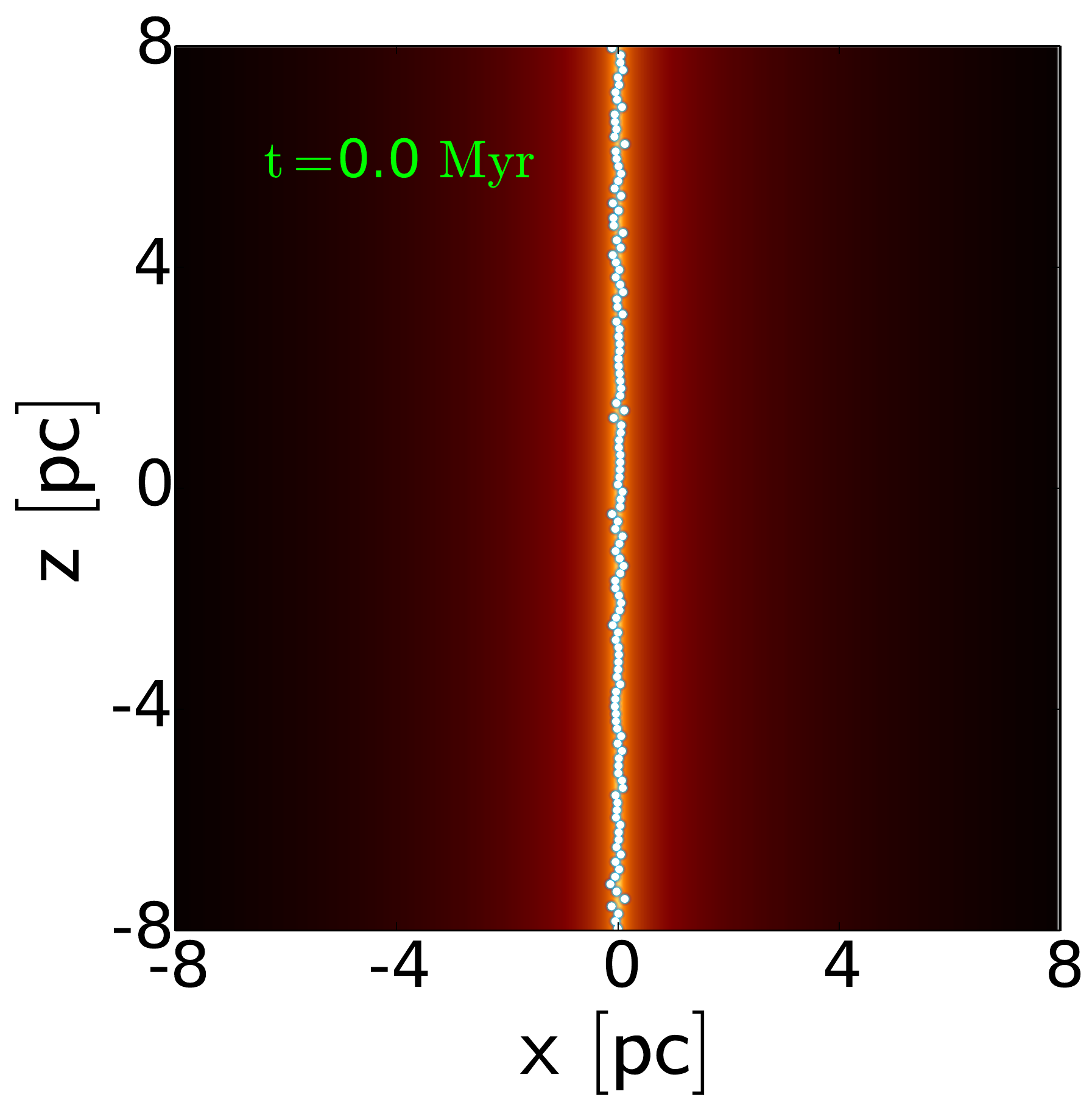} &
\hspace{-4mm}
\includegraphics[height=0.2\textwidth,width=0.2\textwidth]{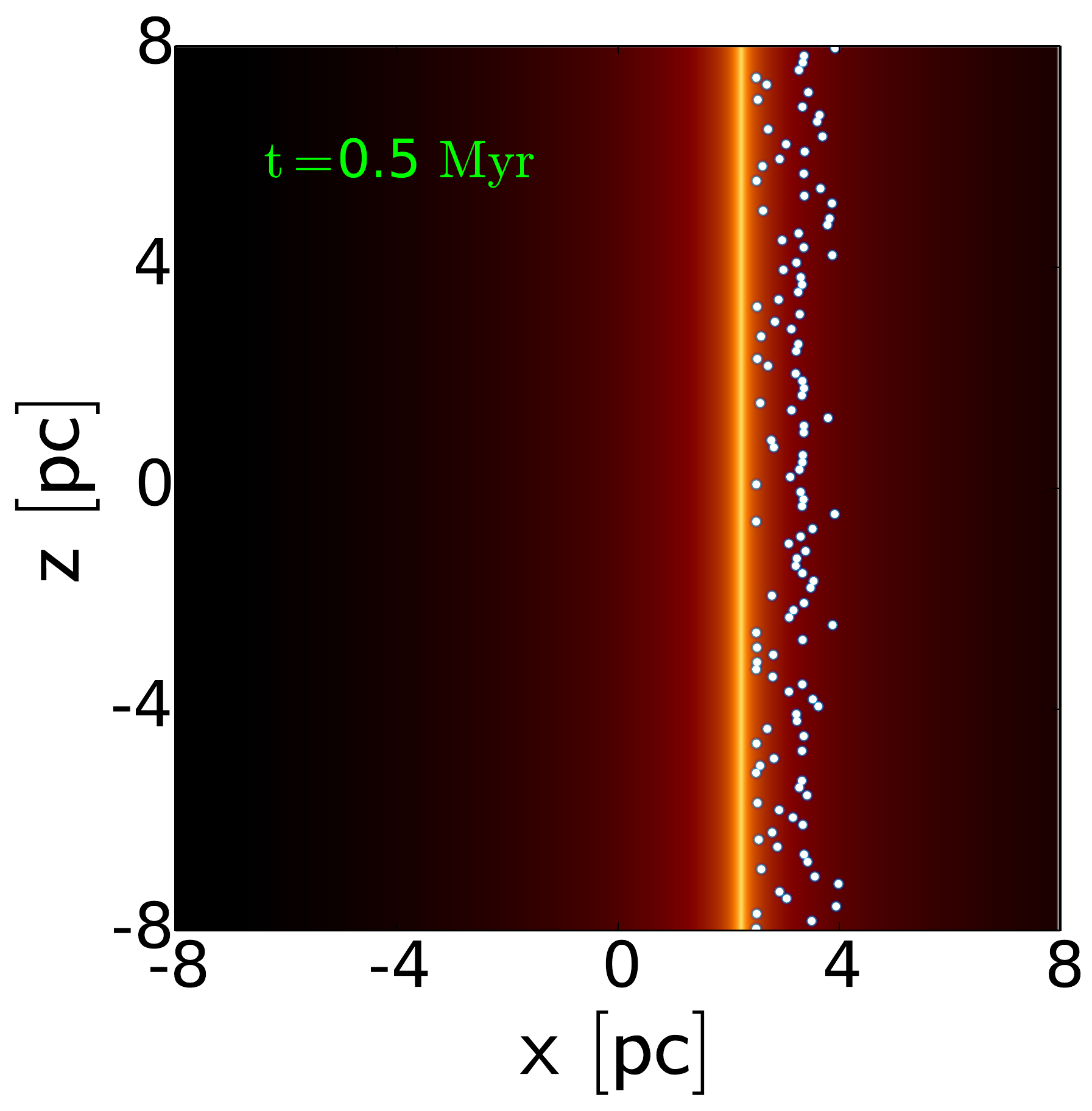}  &
\hspace{-4mm}
\includegraphics[height=0.2\textwidth,width=0.2\textwidth]{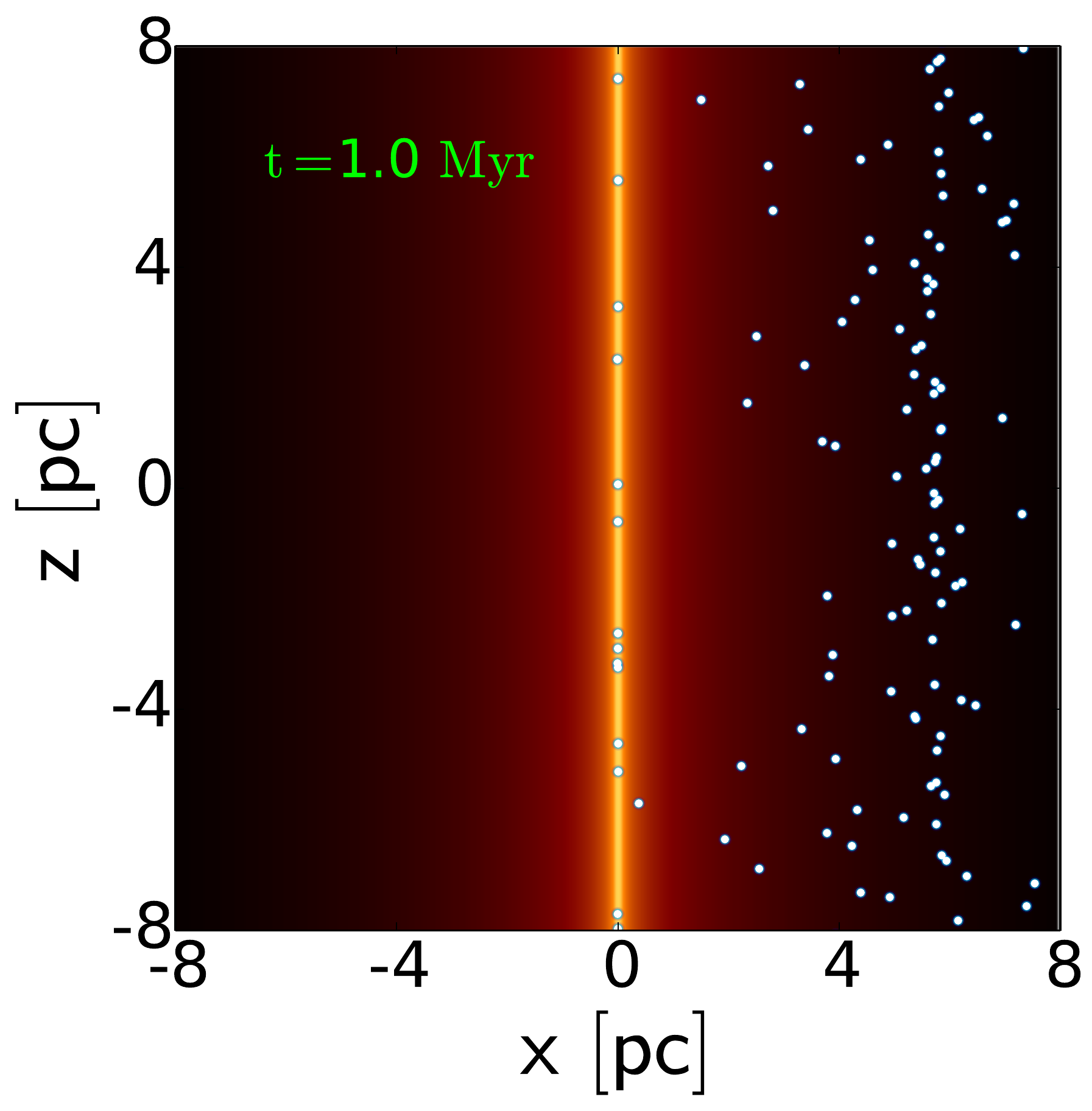} &
\hspace{-4mm}
\includegraphics[height=0.2\textwidth,width=0.2\textwidth]{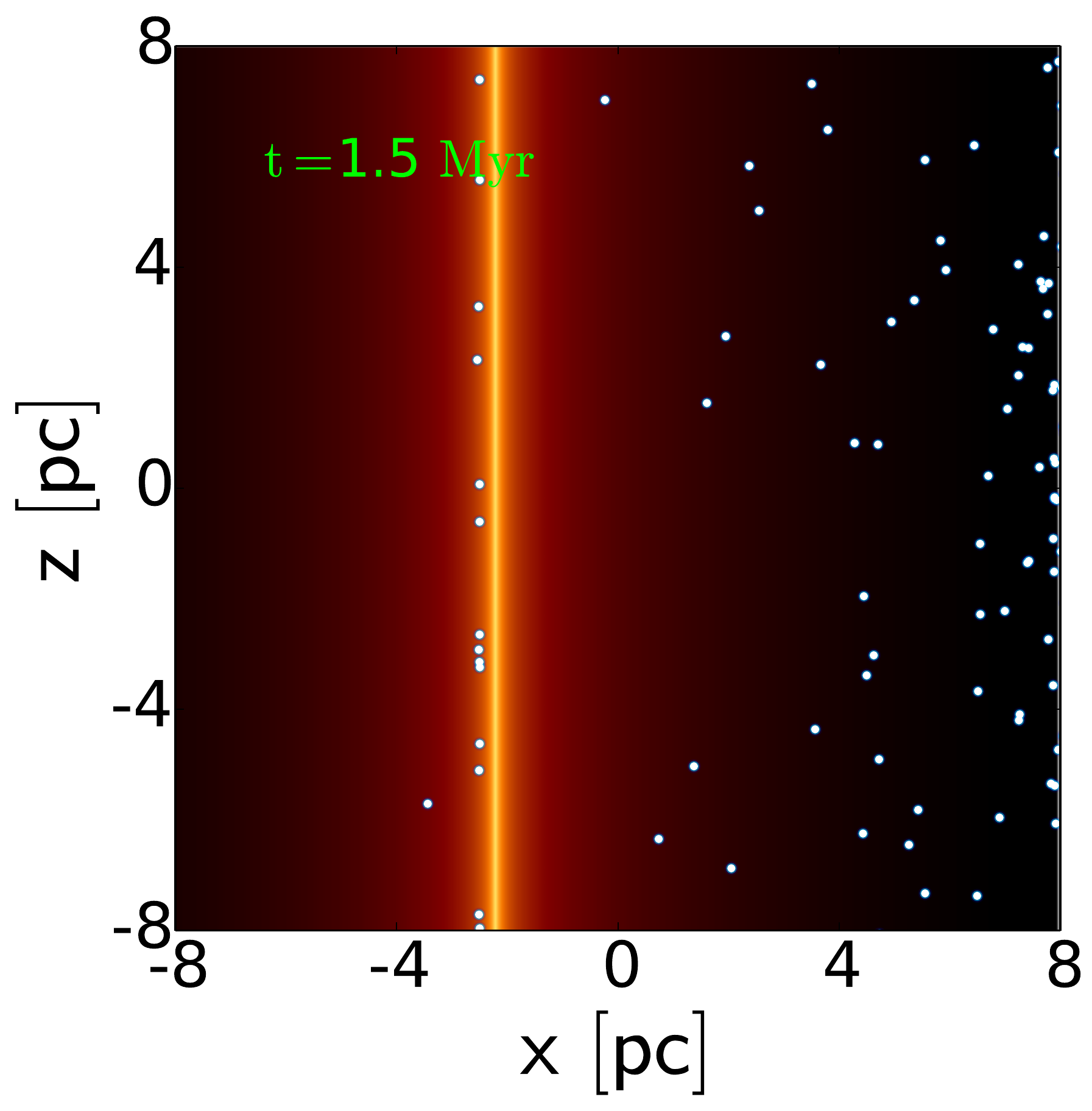} &
\hspace{-4mm}
\includegraphics[height=0.2\textwidth,width=0.2\textwidth]{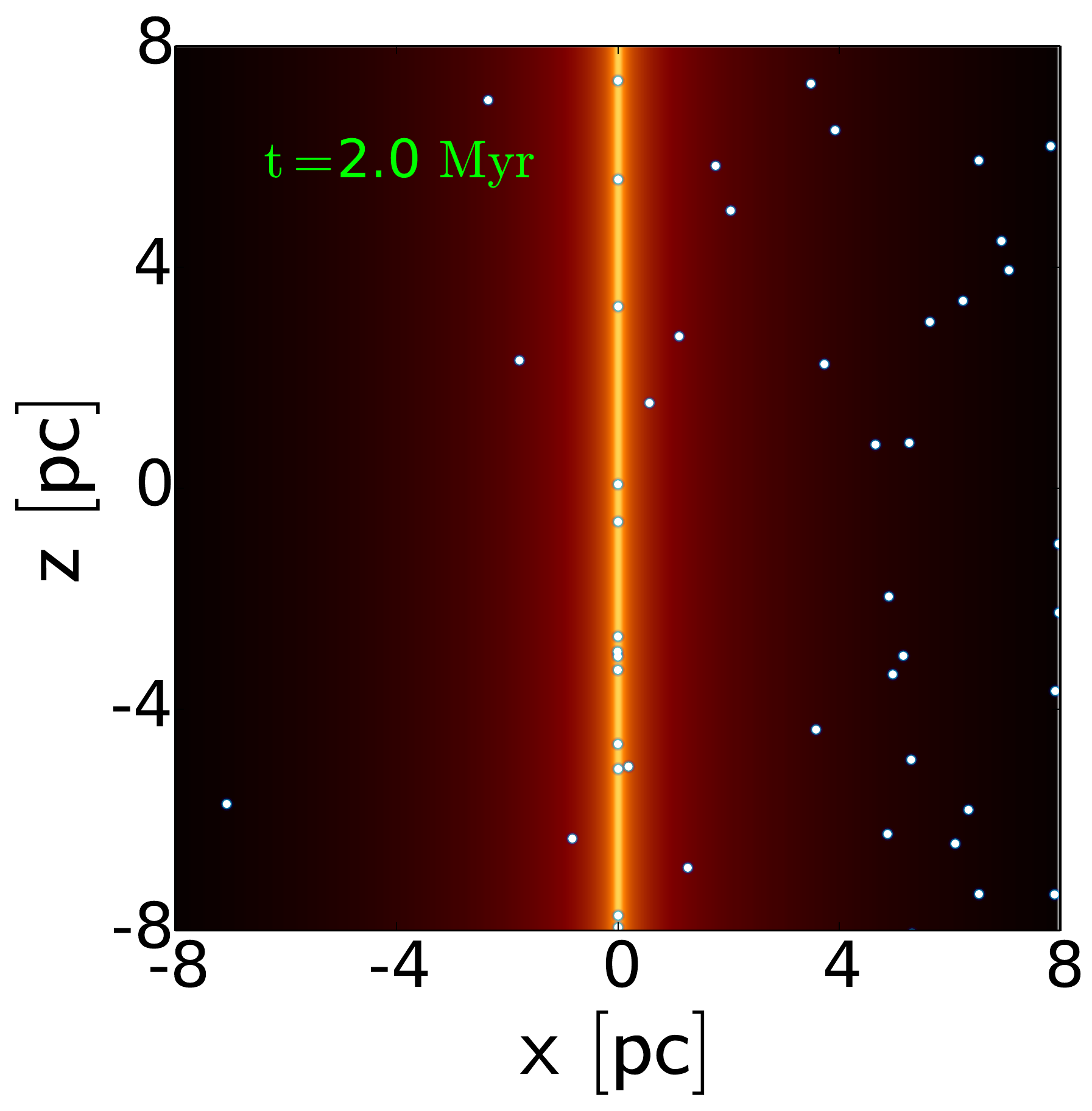}  \\

\end{tabular}  
\caption{Time evolution of a distribution of stars (white dots) in an oscillating filament (red shade represents the density profile). The oscillation period is fixed at P=2\,Myr, while the oscillation amplitude is varied as: $A=0.5$ (top row), $A=1.25$\,pc (middle row) and $A=2.5$ (bottom row). Stars can get ejected from the filament, and the fraction of ejected stars depends on the oscillation parameters of the filament. The simulation for $A=1.25$\,pc results in a symmetric spread of stars around the filament.   }
\label{fig:string_snap}
\end{figure*}

In Fig.~\ref{fig:string_snap} we present three illustrative examples of the effect of an oscillating cylindrical potential on a stellar distribution. We consider one full oscillation of the filament with $P=2$\,Myr, and vary $A$ as 0.5, 1.25, and 2.5\,pc. 

Starting with the simulation with the smallest amplitude, the maximum velocity of the filament is $v_{\rm{max}} = 1.6$\,$\rm{km\,s^{-1}}$. At this relatively low velocity we observe that the stellar distribution remains mostly unaffected by the filament's motion. We also do not observe an increased spread of stars due to stellar interactions. Stellar interactions are negligible as the trajectories are predominately determined by the gas potential. This shows that an ensemble of single stars in a static or slowly moving filament cannot dynamically broaden its distribution by stellar interactions, as we discuss in more detail in Sec.~\ref{sec:discussion:ejec}. 

For the largest amplitude we observe that most stars are immediately ejected after the first turnaround moment of the filament. Initially the stars are dragged along by the gas filament, but as the filament decelerates and then accelerates in the opposite direction, almost all stars will have gained sufficient inertia to leave the filament.   

In the intermediate case, some stars are ejected towards one side of the filament during the first turnaround moment, while a similar fraction of stars is ejected towards the other side at the second turnaround moment. In this particular example, this results in a symmetric spread of stars around the filament. These examples confirm  that an initially narrow distribution of young stars can be broadened due to the filament, under the constraints mentioned in Sec.~\ref{sec:intro}, i.e. within a time scale of 2\,Myr and in a system that is gas dominated.  

\subsection{Filament acceleration and exponential decay}\label{sec:results:expdecay}

\begin{figure*}
\centering
\begin{tabular}{ccc}
\includegraphics[height=0.26\textwidth,width=0.32\textwidth]{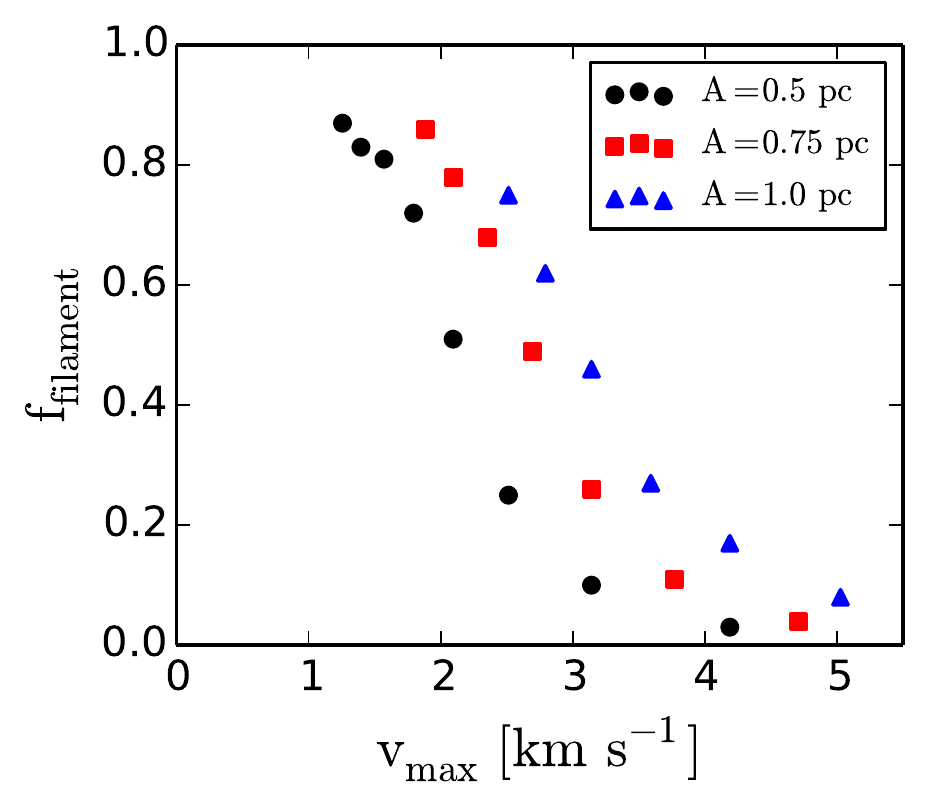} &
\hspace{-4mm}
\includegraphics[height=0.26\textwidth,width=0.32\textwidth]{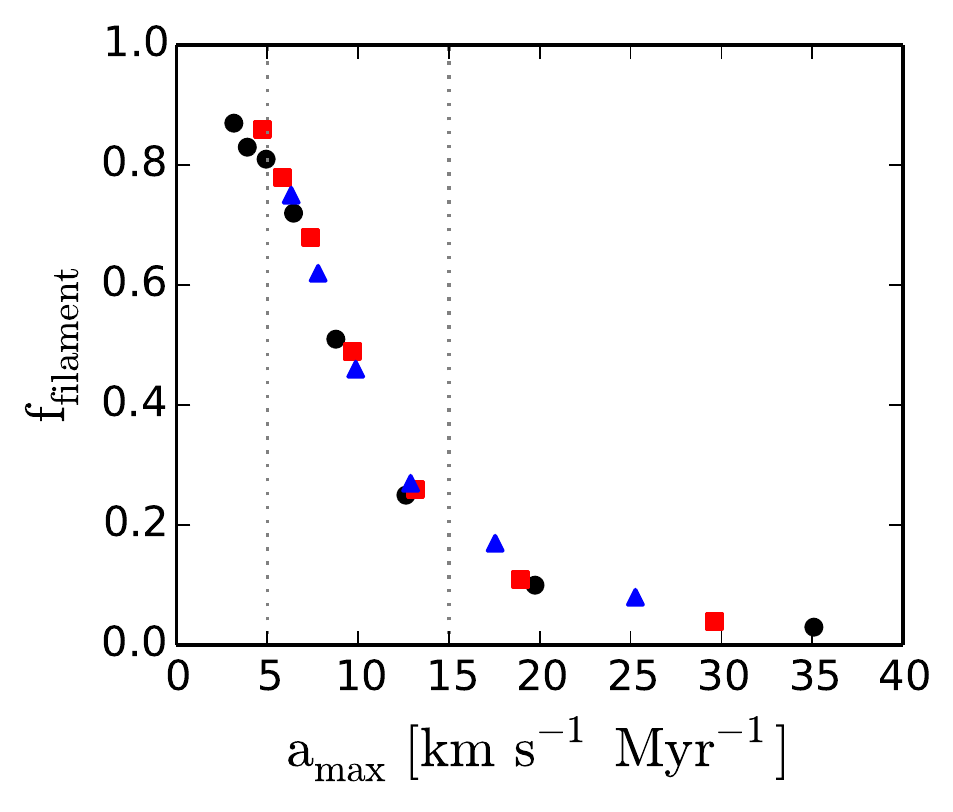} &
\hspace{-4mm}
\includegraphics[height=0.26\textwidth,width=0.32\textwidth]{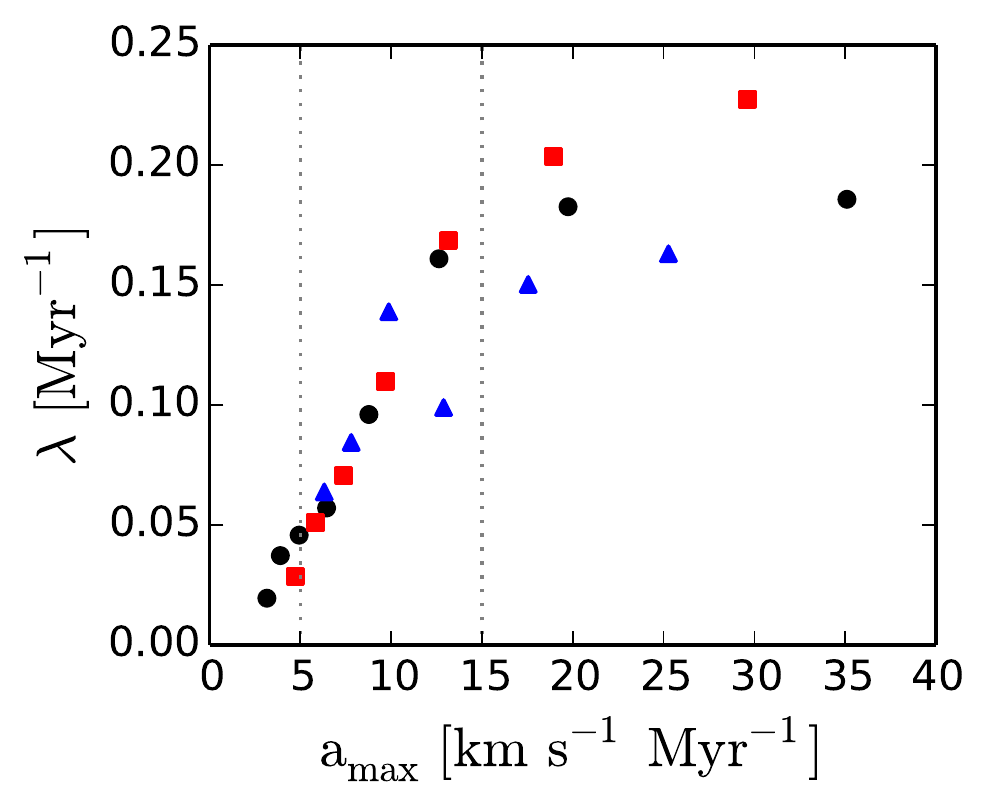}  \\
\end{tabular}  
\caption{ Ejection rate of stars from the filament. In the left panel we plot the fraction of stars still on the filament after 5\,Myr, as a function of maximum velocity of the filament. The different sequences represent different fixed oscillation amplitudes with varying periods. For a given $v_{\rm{max}}$, each sequence gives a different $f_{\rm{filament}}$. In the middle panel we plot $f_{\rm{filament}}$ as a function of the maximum acceleration of the filament, and find that the sequences lie on top of each other. This shows that the ejection rate is purely determined by the acceleration of the filament and not its velocity. In the right panel we plot the fitted exponent of the approximate exponential decay of stars on the filament, $\lambda$, as a function of $a_{\rm{max}}$. The rate of exponential decay is determined by the acceleration of the filament. Only at very large accelerations do most stars immediately escape at the first turnaround of the filament. The data in these diagrams can be found in Tab.~\ref{tab:simulations}. }
\label{fig:exp_decay}
\end{figure*}

We increase the simulation time to 5\,Myr so that we can measure the ejection rate of stars over multiple oscillations of the filament. { Because our model for the gas filament does not provide an escape velocity in the usual sense, we instead adopt a practical "ejection" criterion. We consider a star escaped if its distance to the filament has become significantly larger than the initial spread of the stars: $\Delta r > 5 \sigma_r$. } 
In Fig.~\ref{fig:exp_decay} (left panel) we plot the fraction of stars still on the filament after 5\,Myr, $f_{\rm{filament}}$, as a function of the maximum velocity of the filament, $v_{\rm{max}}$. 
{ We consider a star to be on the filament if it has not escaped the filament according to the criterion defined above, throughout the simulation.} We do this for three sequences of $P$ where each sequence has a different fixed $A$. For a given value of $v_{\rm{max}}$, each sequence has a different maximum acceleration, and we observe that each sequence gives a different value of $f_{\rm{filament}}$. In the middle panel of Fig.~\ref{fig:exp_decay} we plot $f_{\rm{filament}}$ as a function of the maximum acceleration, $a_{\rm{max}}$. For a given value of $a_{\rm{max}}$ each sequence has a different value of $v_{\rm{max}}$, but we find that the different sequences lie on top of each other. This shows that the ejection rate is not determined by the velocity of the filament, but rather by the filament's acceleration. The stellar heating is most effective for $a_{\rm{max}} = 5-15\,\rm{km\,s^{-1}\,Myr^{-1}}$ as in this regime not too many or too few stars are lost from the filament.  

An approximate model for the mass loss rate of stars on the filament can be constructed as follows. If we assume that during each turnaround moment of the filament, a constant fraction, $f$, of stars is ejected, then this will naturally lead to an exponential decay. We fit an exponential to the fraction of stars still on the filament as a function of time and we plot the resultant exponent $\lambda$ as a function of $a_{\rm{max}}$ in Fig.~\ref{fig:exp_decay} (right panel). For small to intermediate accelerations, we observe a tight relation between the exponent and the maximum acceleration of the filament. At large $a_{\rm{max}}$ we observe a deviation from this relation, because the majority of stars is ejected during the first turnaround moment, as illustrated in Fig.~\ref{fig:string_snap} (bottom panels).  

\subsection{Symmetry of the stellar distribution}\label{sec:results:sym}

\begin{figure*}
\centering
\begin{tabular}{cc}
\includegraphics[height=0.29\textwidth,width=0.36\textwidth]{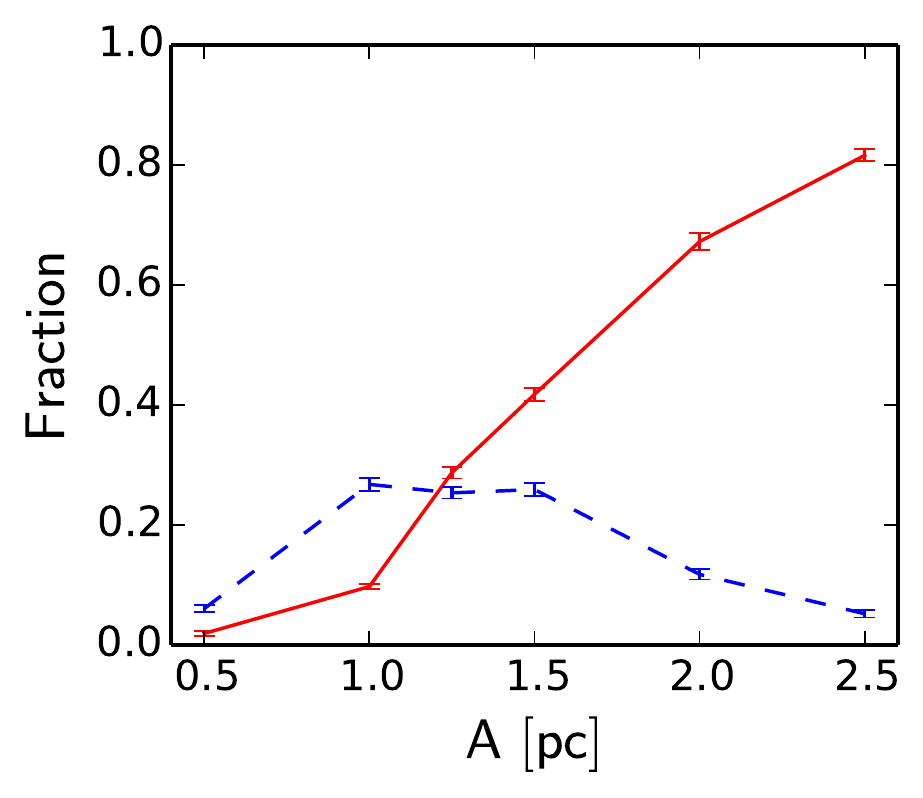} &
\includegraphics[height=0.29\textwidth,width=0.36\textwidth]{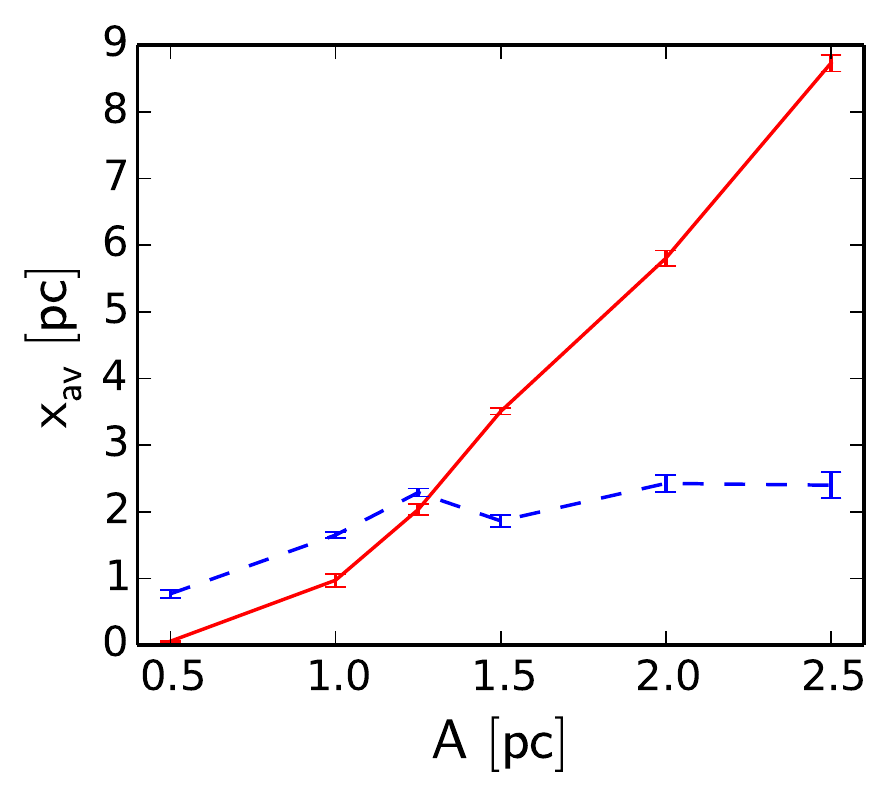}  \\
\includegraphics[height=0.29\textwidth,width=0.36\textwidth]{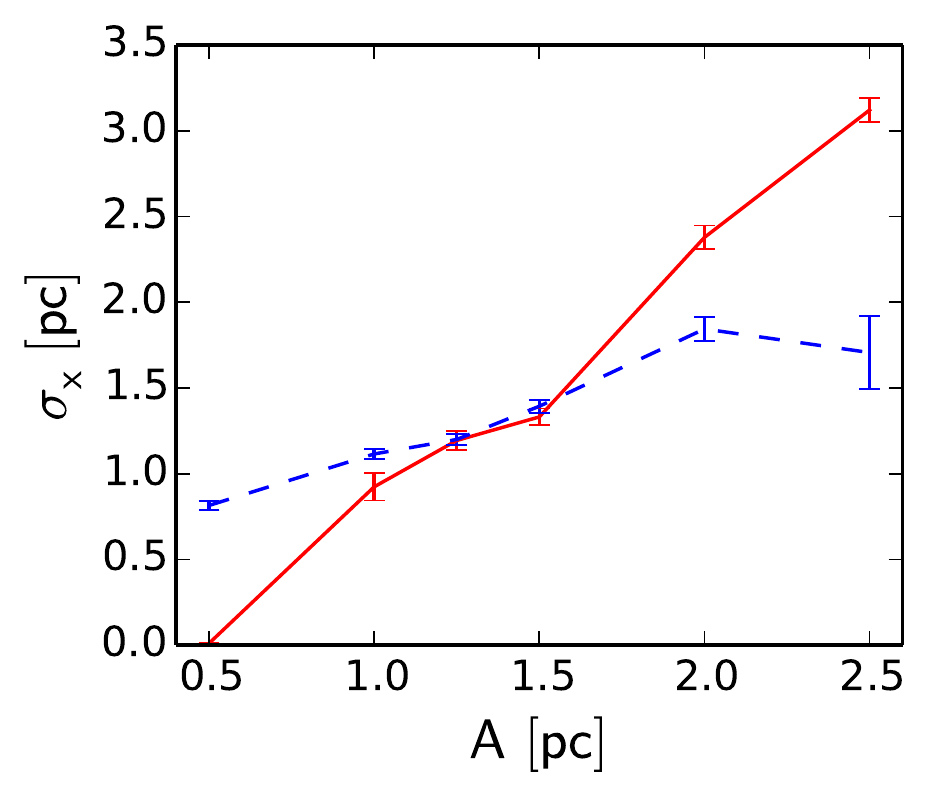}  &
\includegraphics[height=0.29\textwidth,width=0.36\textwidth]{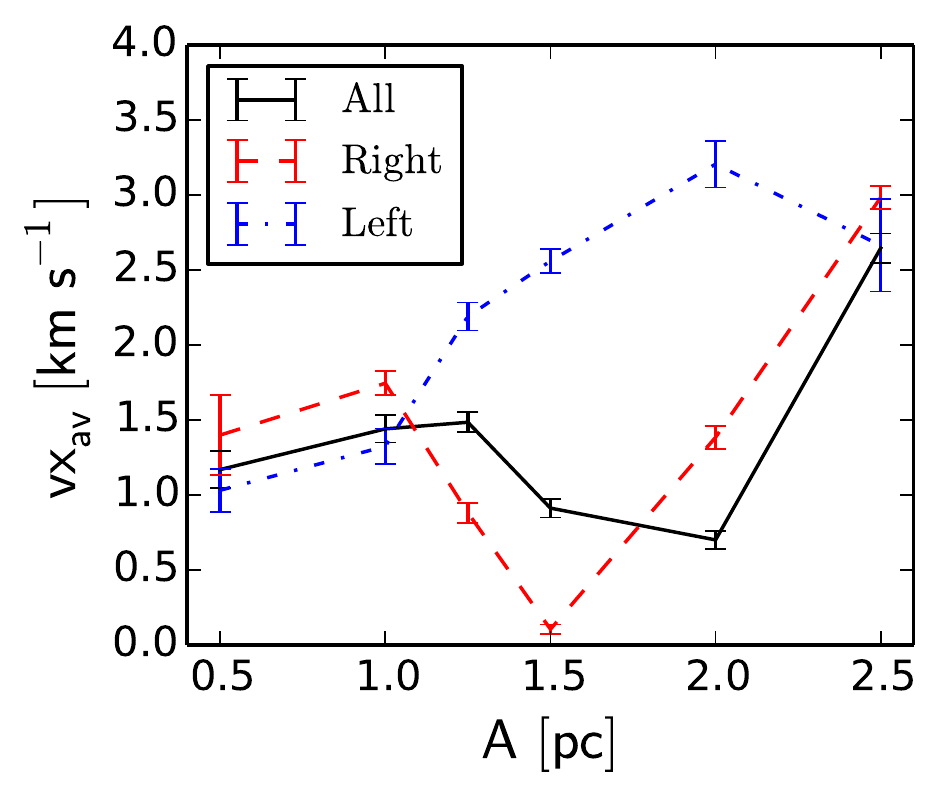}  \\
\end{tabular}  
\caption{ Characterizing the level of symmetry of the stellar distribution around the filament: fraction of stars on each side of the filament (top left), average distance to the filament (top right), stellar spread (bottom left) and average relative velocity to the filament (bottom right). These quantities are plotted separately for the ensemble of stars to the right of the filament (red, solid) and for the ensemble to the left (blue, dashed). We fix the oscillation period to $P=2$\,Myr and vary the oscillation amplitude $A$. We evaluate the systems after one complete oscillation. An approximate symmetric spread is achieved for $A=1.25$\,pc, but the average velocity shows a net bulk motion. The data in these diagrams can be found in Tab.~\ref{tab:simulations2}. }
\label{fig:symmetry}
\end{figure*}

Each simulation performed so far started out with a narrow symmetric distribution of stars on the filament. It is clear that after the first turnaround moment of the filament, the distribution becomes highly asymmetric as stars are only ejected towards one side of the filament. Therefore at least one full oscillation is required in order to be able to obtain a somewhat symmetric distribution, such as observed by \citet{SG16}. 

We consider simulations of one full oscillation of the filament with $P=2$\,Myr and with varying values of $A$. Multiple oscillations would lead to increasingly high values of $a_{\rm{max}}$, unless $A$ is made much smaller than the observed estimate of $A = 1.5$\,pc \citep{SG16}. We use a set of simple quantities that characterize the level of symmetry in the spread of stars around the filament at the end of each simulation. Specifically, we separate the ensemble of stars towards the left of the filament from those towards the right, and measure for each the fraction of stars from the total, the average distance to the filament, and the amount of spread in the distances. We plot the results in Fig.~\ref{fig:symmetry}. In the top left panel, we observe that for small values of $A$, there are more stars towards the left of the filament while for larger values of $A$ most stars can be found on the right. We already observed that for large accelerations most stars are ejected during the first turnaround moment, which explains the behaviour for large $A$. For small $A$ hardly any stars are ejected during the first turnaround moment. Instead, the radial displacements are somewhat increased so that during second turnaround a fraction manages to escape. For $A=1.25$\,pc we observe a crossing point with the same number of stars on both sides of the filament. A similar behaviour can be observed in the top right and bottom left panels of Fig.~\ref{fig:symmetry}, showing the mean distance and stellar spread. These quantities also show a transition point around $A=1.25$\,pc, corresponding to $a_{\rm{max}} = 12.3\,\rm{km\,s^{-1}\,Myr^{-1}}$. This example illustrates that given certain oscillation parameters, $A$ and $P$, a symmetric distribution of stars around the filament is naturally obtained. . 

However, if we inspect the bottom right panel of Fig.~\ref{fig:symmetry}, we observe that the velocity distribution is not symmetric around zero for $A=1.25$\,pc. The solid, black curve representing the total ensemble of stars around the filament, lies roughly halfway between the red and blue curves, reflecting the same number of stars on each side of the filament. The average velocity offset is about 1.5\,$\rm{km\,s^{-1}}$. This can be explained by the fact that the stars that were ejected during first turnaround have had the time to reverse their direction and move back towards the filament. This produces a net bulk motion of the stars with respect to the filament.   

\section{Discussion}\label{sec:discussion}

One of the main novelties of this study is the accelerating, cylindrical background potential. Several dynamical features appear which are not present in for example spherical or disk potentials \citep{BT87}. One of the interesting features in a static cylindrical potential is the possibility of streaming motions along the length of the filament. The only net force is in the radial direction and therefore the stellar trajectories in a gas dominated system are typically a superposition of a radial oscillation and a streaming motion along the filament. { For an oscillating filament, the stars will experience periodic perturbations with varying strengths depending on the orbital phase, which can result in both regular and chaotic behaviour \citep{Halley16} }.

\subsection{On stellar dynamical ejections}\label{sec:discussion:ejec}

We have shown in Fig.~\ref{fig:string_snap} that a narrow distribution of stars in a slowly moving, gas dominated filament cannot spread out radially as it is observed in the ISF in Orion A. 
As mentioned in Sec.~\ref{sec:intro}, some kind of energy source is required. It is well known from star cluster dynamics that binary stars can act as an energy source \citep{Heggie75}. A single star passing a "hard" binary will effectively gain kinetic energy and in principle be able to reach larger distances. 
A simple analytical calculation of the energetics of stellar encounters in a gas filament however, shows that the binary scenario is very unlikely. 

In order for stellar interactions to become more important than the interaction of the stars with the ISF, very small separations are required between the interacting stars. A rough estimate of the separation can be obtained by equating the gravitational potential energy between two stars to the difference in potential energy of two stars in the gas potential:

\begin{equation}
\Delta r \sim \left( {G m \over \phi_0} \right)^{{1 \over \gamma + 1 }},
\label{eq:1}
\end{equation}   

\noindent with $\Delta r$ the separation, $G$ the gravitational constant, $m$ the mass of a star, and $\phi_0$ and $\gamma$ the two parameters describing the power law profile of the gas potential \citep[Eq.~8]{SG16}. Filling in $m = 0.5\,\rm{M_\odot}$ as the typical mass of a young stellar object, $\phi_0 = 6.3\,\rm{(km\,s^{-1})}^2$ and $\gamma = 3/8$ \citep{SG16}, we obtain a separation of about 600\,AU. 

Another constraint is that the ejected stars have to climb out of the potential well of the filament. As we have shown in Sec.~\ref{Sec:method:validation}, a star at the centre of the filament requires a velocity of 3\,$\rm{km\,s^{-1}}$ to reach a distance of 1\,pc. In order to estimate the initial separation of the stars leading to such an ejection velocity, we consider a chaotic three-body interaction in the centre of the gas filament. The typical kinetic energy of the ejected single star can be estimated from \citet[Fig.~5]{Brutus} and is about $EK \sim 0.4\,E$, with $EK$ the kinetic energy of the ejected star and $E$ the absolute total energy of the three-body configuration. Using the appropriate expressions for the kinetic energy and total energy of an initially cold system, we estimate:

\begin{equation}
\Delta r \sim 0.4 G {M^2 \over m v^2}, 
\label{eq:2}
\end{equation}

\noindent with $M$ the total mass, $m$ the mass of a single star and $v$ the ejection velocity. Filling in the numbers given above we obtain a separation of about 170\,AU. Another way to estimate the maximum separation is by considering a binary-single star encounter in which the resultant ejected star has a velocity of order the original orbital velocity of the binary, e.g. $v \sim \sqrt{GM/r}$, resulting in a separation of about 80\,AU.  

In summary, maximum separations of order 200\,AU are required in order to be stellar mass dominated and to be able to produce the required ejection velocities. Observations of young stellar objects on the ISF show that the initial separation 
{ distribution peaks between 5 and 15 kAU \citep[][Fig.~7]{separation16}. }
Also, the multiplicity of young stellar objects on the ISF is observed to be about 12-14\% for separations of 100-1000 AU (Kounkel et al. 2016b, Fig.~13)\nocite{Kounkelb}, and only 5\% for separations smaller than about 10\,AU (Kounkel et al. 2016a)\nocite{Kounkela}.
Based on the discrepancy between the observations and analytical estimates presented in this section, we conclude that the stellar interactions cannot be the main heating mechanism of the stars, but could in principle make a minor contribution to the broadening of the stellar distribution. 

This conclusion could not have been guessed from pure N-body simulations, i.e. simulations that ignore the gaseous component \citep[e.g.][]{Parker17}. The presence of a significant gas mass will generally suppress stellar interactions and at the same time increase the potential well out of which the stars have to escape. 
Accounting for the gas mass distribution, that is, the gravitational potential (e.g. Smith et al. 2011a and 2011b, Farias et al. 2015, and Stutz \& Gould 2016)\nocite{Smith11a, Smith11b, Farias15, SG16} is an absolute requirement for the accurate modelling of the kinematics and dynamics of stars in star forming regions, for which observations are becoming increasingly routine \citep{Tobin09, Foster15, riglia16, Sacco17, dario17}. This approach is essential for understanding relevant aspects of the star formation process as well as the  evolution of stellar systems after gas dispersal \citep[e.g.][]{Smith13}.

\subsection{The slingshot mechanism}

This study shows that, in principle, an accelerating filament can provide a mechanism to dynamically heat up the stars within the required time scale, as long as we are in the regime where the slingshot mechanism is most effective, i.e. filament accelerations of order $\rm{a_{\rm{max}} = 5-15\,\rm{km\,s^{-1}\,Myr^{-1}}}$ (see Sec.~\ref{sec:results:expdecay}). 

In this first exploratory study, we assume a simple sinusoidal motion of the filament. This model predicts that if the spread of stars around the filament is symmetric, then the stars should show a bulk motion with respect to the filament (see Sec.~\ref{sec:results:sym}). Our estimate of $\rm{1.5\,km\,s^{-1}}$ is larger than the observed value of 0.45\,$\rm{km\,s^{-1}}$ for young stellar objects around the ISF \citep{SG16}, but we have neglected any projection effects that reduce our estimate, and in reality, the filament motion may be more complex than what we have presented in this study.  

Our results here show that dynamically heating up star clusters through oscillating filaments is a potential mechanism to explain the kinematical data of the gas and the stars in the ISF. While the results presented here are very promising, future research directions on the one hand include further explorations of the filamentary motion itself, taking into account both the magnetic tension force as well as gravitational interactions, but also the interaction of the filament with more complex structures, such as for instance a small stellar cluster.
Independent time scale estimates for the gas formation and evolution via for example chemical means \citep[e.g.][]{Koertgen17}, as well as the determination of magnetic field properties \citep[e.g.][]{Busquet13, Pillai16, Reissl17} are essential. 

\section{Conclusion}\label{sec:conclusion}

We present the first numerical investigation of the dynamical evolution of young stars in an accelerating, cylindrical potential. Such a configuration is an accurate description of the integral shaped filament (ISF) in Orion A, in which young stars interact with the gas filament \citep{SG16}. We model the ISF with a cylindrical potential and allow it to oscillate in a sinusoidal fashion, in a direction perpendicular to the length of the filament. By coupling N-body dynamics to this background potential, we are able to simulate the dynamics and kinematics of young stars in filamentary systems. 

Our results show that the slingshot mechanism \citep{SG16} can successfully reproduce several observed features in the stellar distribution around the ISF:
\begin{itemize}
\item An initially narrow distribution of stars along the central ridgeline of the filament can be broadened due to the filament's motion. This is accomplished in a time scale of about 2-3\,Myr and with the observed spread of a few parsecs (see beginning of Sec.\ref{sec:results}).
\item The stellar heating is caused by the acceleration of the filament, and the regime in which the slingshot mechanism works most effectively is about $a_{\rm{max}}\rm{ = 5-15\,\rm{km\,s^{-1}\,Myr^{-1}}}$ (see Sec.\ref{sec:results:expdecay}).
\item A symmetric spread of stars around the filament is obtained naturally by ejecting a fraction of stars on both sides of the filament during at least one complete oscillation of the filament (see Sec.~\ref{sec:results:sym}). 
\item If the positional spread is symmetric, our model predicts a bulk motion of the stars with respect to the filament, e.g. there is a non-zero average velocity with respect to the filament's motion (see last paragraph of Sec.~\ref{sec:results:sym}).
\item Stellar interactions are not the dominant mechanism in the ISF as the required stellar density and multiplicity is much higher than what is observed (see Sec.~\ref{sec:discussion:ejec}). 
\end{itemize}

These conclusions motivate further research on the slingshot mechanism. Research directions include the formation and evolution of star clusters in accelerating, cylindrical potentials and the formation and evolution of protostars in dense gas filaments.  

\section{Acknowledgements}

AS and TB thank Andrew Gould for many stimulating discussions. 
AS, DRGS, MF and TB are thankful for funding from the ''Concurso Proyectos Internacionales de Investigaci\'on, Convocatoria 2015'' (project code PII20150171) and the BASAL Centro de Astrof\'isica y Tecnolog\'ias Afines (CATA) PFB-06/2007. DRGS acknowledges funding through Fondecyt regular (project code 1161247) and ALMA-Conicyt (project code 31160001).
TB acknowledges support from Fundaç\~ ao para a Ci\^ encia e a Tecnologia in the form of the grant SFRH/BPD/122325/2016 and from CIDMA strategic project UID/MAT/04106/2013.

\bibliographystyle{mn2e}      
\bibliography{numerical_slingshot_arxiv}      

\end{document}